\RequirePackage[2018-12-01]{latexrelease}
\documentclass{gGAF2e}
\usepackage{graphicx}
\usepackage{xcolor}
\usepackage{bm}
\usepackage{mathrsfs}
\usepackage{amsfonts}

\newsavebox{\astrutbox}
\sbox{\astrutbox}{\rule[-5pt]{0pt}{20pt}}

\newcommand\beq{\begin{equation}}
\newcommand\eeq{\end{equation}}
\newcommand\beqa{\begin{eqnarray}}
\newcommand\eeqa{\end{eqnarray}}

\newcommand{\bnabla}{\mbox{\boldmath$\nabla$}}

\newcommand{\eps}{\varepsilon}

\def\bnab{\mbox{\boldmath $ \nabla$}}

\newcommand{\bR}{\mathbf{\hat{r}}}
\newcommand{\bx}{\mathbf{x}}

\newcommand{\bhz}{\mathbf{\hat{z}}}
\newcommand{\bhphi}{\mbox{\boldmath $\hat{\phi}$}}
\newcommand{\bhs}{\mathbf{\hat{s}}}

\newcommand{\bu}{\mathbf{u}}
\newcommand{\btu}{{\mathbf{\tilde{u}}}}

\def\<{\langle}
\def\>{\rangle}

\newcommand{\lam}{\lambda}

\newcommand{\V}{{\mathscr V}}
\newcommand{\E}{{\epsilon}}
\newcommand{\UU}{\check{U}}
\newcommand{\VV}{\check{V}}
\newcommand{\WW}{\check{W}}
\newcommand{\UUU}{\check{\check{U}}}
\newcommand{\VVV}{\check{\check{V}}}
\newcommand{\WWW}{\check{\check{W}}}
\newcommand{\bUU}{{\bf \check{U}}}

\newcommand{\TT}{\mathcal{T}}
\newcommand{\PP}{\mathcal{P}}

\begin{document}




\title{Weakening the effect of boundaries: `diffusion-free' boundary conditions as a `do least harm' alternative to Neumann}

\author{Yufeng Lin${\dag}$ and Rich R. Kerswell${\ddag}$ $^{\ast}$\thanks{$^\ast$
\vspace{6pt} Corresponding author. Email: rrk26@cam.ac.uk}
\\\vspace{6pt}  ${\dag}$ Department of Earth and Space Sciences, Southern University of Science and Technology, Shenzhen 518055, P.R. China\\ ${\ddag}$DAMTP, Centre for Mathematical Sciences, University of Cambridge, CB3 0WA, UK
\\\vspace{6pt}\received{\today} }

\maketitle

\begin{abstract}
In this note,  we discuss a poorly known alternative boundary condition to the usual Neumann or `stress-free' boundary condition typically used to weaken boundary layers when diffusion is present but very small.  These `diffusion-free' boundary conditions were first developed (as far as the authors know) in 1995 (Sureshkumar \& Beris {\em J.  Non-Newtonian Fluid Mech.} {\bf 60}, 53-80, 1995)  in viscoelastic flow modelling but are worthy of general consideration in other research areas. To illustrate their use,  we solve two simple ODE problems and then treat a PDE problem - the inertial wave eigenvalue problem in a rotating cylinder, sphere and spherical shell for small but non-zero Ekman number $E$. Where inviscid inertial waves exist (cylinder and sphere), the viscous flows in the Ekman boundary layer are $O(E^{1/2})$ weaker than for the corresponding stress-free layer and fully $O(E)$ weaker than in a non-slip layer. These diffusion-free boundary conditions can also be used with hyperdiffusion and provide a systematic way to generate as many further boundary conditions as required. The weakening effect of this boundary condition could allow precious numerical resources to focus on other areas of the flow and thereby make smaller,  more realistic values of diffusion accessible to simulations.
\end{abstract}


\section{Introduction}
In many flow situations of interest,  diffusion of a given physical quantity may be so small relative to other effects that it would appear  safely neglectable. The well-known caveat to this is the presence of boundaries where small scales can make diffusion a leading order effect to accommodate 
the boundary conditions. The situation can be even more complicated if these  boundary layers also generate internal shear layers in which  
diffusion plays an enhanced role  in the bulk (e.g. the inertial wave problem to be discussed below). The strategy almost always  is either to take the smallest diffusion coefficient possible which still allows accurate solutions to be computed (even though this may be orders of magnitude bigger than the actual value) and/or the boundary conditions are adjusted to weaken the boundary layers and possibly internal layers which result. In the case of high Reynolds number flows (or low Ekman numbers) where momentum diffusion is small, the latter is invariably achieved by adopting `stress-free' boundary conditions instead of the usual `non-slip' conditions of the flow at a solid boundary \citep[e.g.][]{Vidal23}. This, however, can create its own problems in that stress-free boundary conditions prevent any angular momentum transfer through viscous stresses in rotating systems \citep{Vidal23}.

The purpose of this short note is to describe a further option - hereafter referred to as `diffusion-free' boundary conditions - for mitigating the effects of diffusion and to demonstrate that they produce an even weaker viscous boundary layer correction than `stress-free' or Neumann conditions. In the interests of clarity, we specialise below to the vanishing viscosity situation although it should be clear that what follows is perfectly general for handling the diffusion of any physical quantity, e.g. a scalar like density, another vector such as a magnetic field or even a tensor like the conformation tensor in viscoelastic fluids. In fact, it is within the viscoelastic flow community that the new boundary condition was (as far as the authors are aware) first used \citep{Sureshkumar95, Sureshkumar97}. Here, the correct boundary condition on the conformation tensor, which describes an ensemble-averaged configuration of the polymers, is unclear and so those authors decided to simply turn the diffusion off at the boundary. This meant that  the diffusion-free equation was imposed at the wall as the required extra boundary conditions or, equivalently, for a $C^2$ solution, the (tensorial) Laplacian of the conformation tensor had to vanish.  Carrying this over to the flow velocity, the idea is to augment the usual (inviscid) no-penetration normal condition with the  extra conditions that the tangential components of the inviscid equation hold there. Again for a sufficiently smooth velocity field, this is equivalent to imposing that the tangential components of the viscous (vectorial) Laplacian term vanish.  Imposing 
that the second derivative must vanish when solving a second order equation appears paradoxical. But, in fact, all this does for a smooth solution is impose that the `rest' of the equation (comprising of first and zeroth derivative terms) vanishes at the boundary so the diffusion-free boundary condition is equivalent to a generalised Robin condition (`generalised' as this condition can be a nonlinear relation between zeroth and first derivatives). 

Boundary conditions which include the second derivative for a second order equation are not new. They have been been studied since the 1950s \citep{Feller52, Ventcel59} in the mathematical analysis literature and variously go  under the names of `Ventcel' (e.g. \cite{Buffe17}), `Ventsell' (e.g. \cite{Hinz18}) or `Wentzell' (e.g. \cite{Denk21}) boundary conditions. The novelty introduced by \cite{Sureshkumar95} was to implicitly use a special form of them  when artificially smoothening a differential equation.

We first illustrate the diffusion-free boundary condition for  a couple  of simple ODE problems and then to demonstrate how it works for a realistic  PDE case.
The inertial wave problem is chosen for the latter as: a) it is relevant for  rapidly rotating systems such as the Earth's core  \citep{Vidal23} where the exact tangential boundary conditions are unclear; b) it has some challenging viscous features to resolve in the form of internal shear layers \citep{McEwan70}; c) it is still an area of active research \citep[e.g.][]{He22, He23, LeDizes23}; and finally d) it is an opportunity to update some 30-year-old work by one of the authors \citep[][hereafter KB95]{KB95}.

\section{\label{first_model}Model Problem 1: Nonlinear Robin condition}
%

We first start by considering the simple, albeit nonlinear, ODE problem
\beq
y'=-y^2
\eeq
over $x\in [0,1]$ ($y':=dy/dx$) with boundary condition $y(0)=1$. This has the solution $y=1/(1+x)$.  Now imagine that there is actually very weak diffusion in the system and really the equation is
\beq
y'=-y^2+\eps y{''}
\label{model_1}
\eeq
with the diffusion coefficient $\eps \ll 1$. The question then arises as to what extra boundary condition to impose at $x=1$ because  the problem is now second order.  If known,  the physically-relevant condition  is the natural one, of course,  but sometimes this is unclear  or there is a need to minimise the effect of the boundary to examine its influence or perhaps reduce the numerical  resolution needed to capture the interior solution. Then there is interest in selecting the boundary condition which does `least harm' to the interior solution. To explore this, we now examine the effect of imposing: (i) a Dirichlet condition $y(1)=0$; (ii) a Neumann condition $y'(1)=0$; or (iii) the new `diffusion-free' condition $y'(1)=-y(1)^2$ which is equivalent to $y''(1)=0$ for a smooth solution.

The three problems can be straightforwardly solved  using matched asymptotic expansions to give the leading composite solutions:
\beq
\begin{array}{lll}
(i)   &  y = \displaystyle{\frac{1}{1+x}
-\frac{1}{2} e^{-(1-x)/\eps}  }         \hspace*{2cm}  & {\rm Dirichlet}, \\
& & \\
(ii)  &  y = \displaystyle{ \frac{1}{1+x} 
+\frac{1}{4}\eps e^{-(1-x)/\eps}   }                  & {\rm Neumann}, \\
& & \\
(iii) &  y =\displaystyle{\frac{1}{1+x}
-\frac{1}{4}\eps^2 e^{-(1-x)/\eps} }                  & {\rm Diffusion\!\!-\!\!free}.
\end{array}
\eeq
%
%
The first term on the RHS represents the leading `outer' solution (with the next correction being  $O(\eps)$) and  each second term is the leading `inner' boundary layer correction. The take-home message here is that this is $O(1)$ for the Dirichlet condition, $O(\eps)$ for the Neumann condition and $O(\eps^2)$ for the diffusion-free condition which therefore creates the weakest boundary layer correction.  It's worth remarking also that the actual condition imposed is $y'(1)=-y(1)^2$ which is not the usual (linear) Robin condition due to the nonlinearity in $y$ but a suitable generalisation i.e. a nonlinear relationship between $y$ and $y'$ which is imposed at $x=1$. 

At this point the reader may wonder about asking for an even higher derivative to vanish at $x=1$ or using a higher order, hyperdiffusion term. The former can work to reduce the boundary correction to be $o(\eps^2)$ but doesn't change  the fact that the diffusion term gives an $O(\eps^2)$ perturbation in the interior anyway. The latter introduces the need for yet more boundary conditions but again diffusion-free b.c.s and its derivatives well work - see Appendix A.

%
%
%
%
%
%
%
%

\section{Model Problem 2: Interior smoothening}
%

The second ODE problem we consider is 

\beq
y=(2+x^2)H(x)\,:= \left\{ \begin{array}{lr}   2+x^2  &  \qquad x \geq 0\\
0      &  \qquad x <0 
\end{array}
\right.
\label{inviscid_model}
\eeq
for $x \in[-1,1]$ and $H(x)$ is the Heaviside function. This has a discontinuity at $x=0$ such as could be present in a hyperbolic system and non-zero first and second derivatives at one boundary for reasons which will become clear immediately below.   To smooth  the solution (e.g. to numerically estimate it),  diffusion could be added as follows
\beq
y = (2+x^2)H(x)+\eps^2 y''
\label{viscous_model}
\eeq
to target the discontinuity hopefully leaving the rest of the solution largely unchanged (the diffusion coefficient $\eps^2$ controls the amount of smoothening). This makes the problem second order requiring two new boundary conditions. We now show again that the effect of  the diffusion-free boundary conditions will be less disruptive on the original solution than adopting Neumann (`stress-free') conditions.

%
%
\begin{figure}
\centering
\scalebox{0.37}[0.37]{\includegraphics{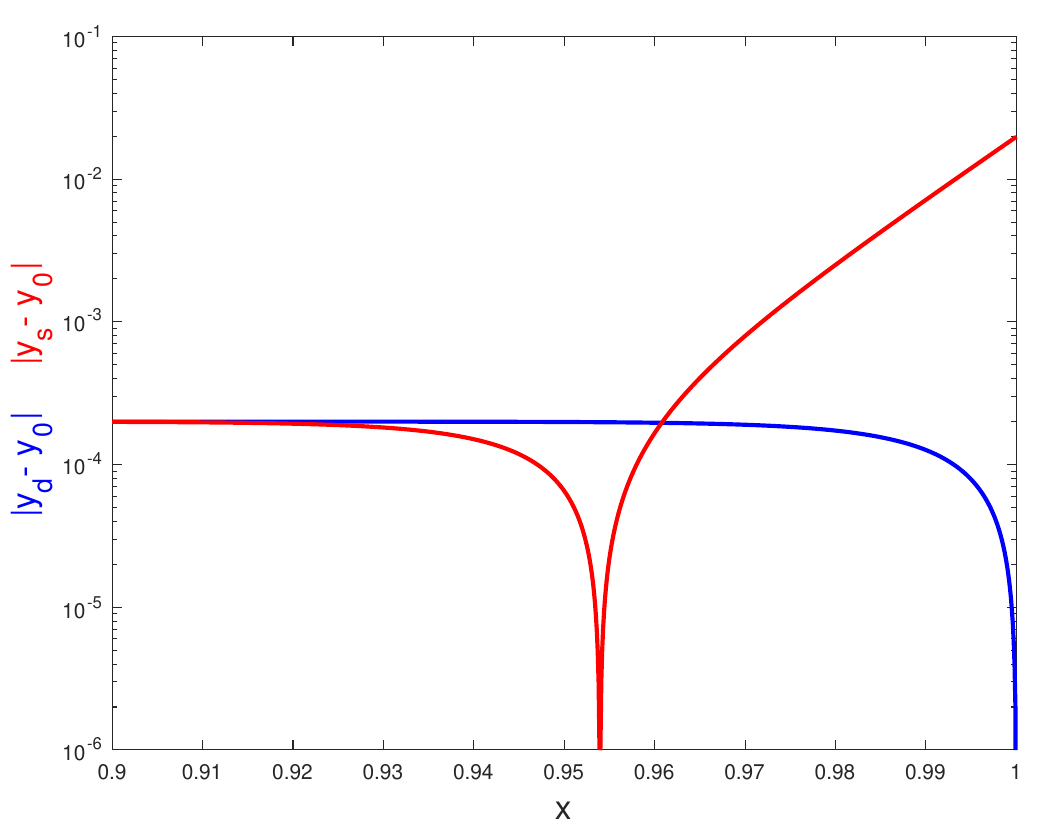}}
\scalebox{0.37}[0.37]{\includegraphics{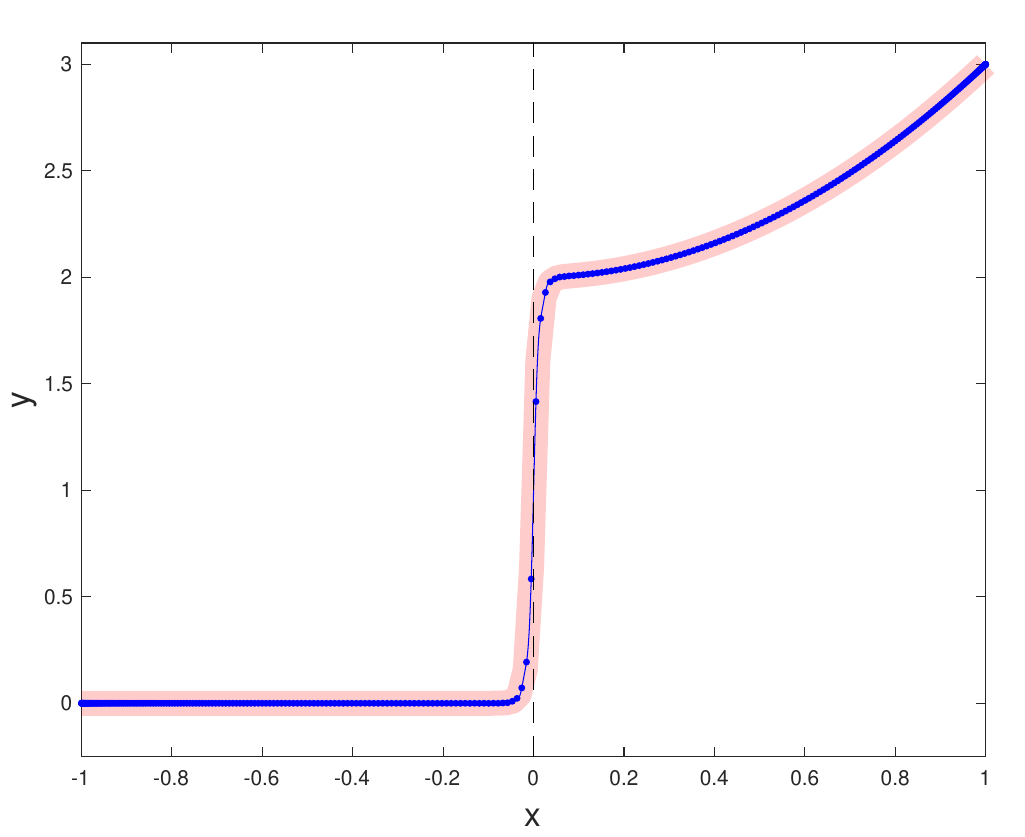}}
\caption{\label{fig1} Left: The difference between the stress-free solution  $y_s$ and the diffusion-free solution $y_0$ is plotted on a log scale in red together with the corresponding difference between the diffusion-free boundary condition solution (\ref{diffusion_free_soln}), $y_d$,  and $y_0$ (in blue) for  $\eps=0.01$ near $x=1$. This plot  demonstrates that the stress-free solution  produces a larger deviation at the boundary in $y$: both have an unavoidable interior deviation of $O(\eps^2)$ due to the change in the ODE and an $O(1)$ deviation at the discontinuity where $x=O(\eps)$.  Right: a numerical solution for  (\ref{diffusion_free_soln}) shown as the thin blue line again for  $\eps=0.01$ using a spectral method with 300 Chebyshev polynomials (blue dots indicate the collocation points used; it is expected that $O(1/\eps)$ polynomials are needed due to sparse collocation point spacing at the middle of  the $[-1,1]$ range). The thick transparent red line is the analytic expression  $y_d(x)$ given in (\ref{diffusion_free_soln}). Note that the same result is obtained imposing the equivalent boundary condition that $y''=0$ at $x=\pm 1$.}
\label{model}
\end{figure}

%
%
\subsection{Neumann boundary conditions}

Neumann (stress-free) boundary conditions are 
\beq
y'(-1)=0 \quad \& \quad y'(1)=0
\eeq
which are often viewed as minimally disruptive. This choice leads to the leading asymptotic solution
\beq
y_{s} \sim \left\{ \begin{array}{cr}      
\displaystyle{2(1+\eps^2) e^{ -1/\eps }\cosh \left(\frac{1+x}{\eps} \right) }& x <0, \\
 \displaystyle{2+x^2+2\eps^2 + 2\eps \sinh \left( \frac{1-x}{\eps}\right) -\biggl[\,2\eps+2(1+\eps^2) e^{ -1/\eps } \,\biggr] \cosh \left( \frac{1-x}{\eps} \right) }& x \geq 0 .
\end{array} \right. 
\label{stress_free_soln}
\eeq
There are then 3 distinct scalings for how this solution differs from the diffusion-free solution $y_0$ of  (\ref{inviscid_model}): for $x=0(\eps)$,  $y_s-y_0 =O(1)$,  for $1-x=O(\eps)$,  $y_s-y_0=O(\eps)$ to fix up the boundary condition,  and otherwise $y_s-y_0=O(\eps^2)$ - see Figure \ref{model}.

\subsection{Diffusion-free boundary conditions}
%
%
We now impose `diffusion-free' boundary conditions   which are  just 
\beq
y(-1)=(2+x^2)H(x)|_{x=-1}=0 \quad \& \quad y(1)=(2+x^2)H(x)|_{x=1}=3
\label{d_free_bcs}
\eeq
The leading asymptotic solution is  then 
\beq
y_{d} \sim \left\{ \begin{array}{cr}     
\displaystyle{ 2(1+\eps^2) e^{ -1/\eps } \sinh \left( \frac{1+x}{\eps} \right) }
& x < 0,\\ 
\displaystyle{2+x^2+ 2\eps^2 \biggl[\,1-\cosh \left( \frac{1-x}{\eps} \right) \, \biggr]
+\biggl[\,2\eps^2-2(1+\eps^2) e^{-1/\eps } \,\biggr] \sinh \left( \frac{1-x}{\eps} \right) }
& x \geq 0. 
\end{array} \right.
\label{diffusion_free_soln}
\eeq
There are now only 2  distinct scalings for how this solution differs from the solution $y_0$ of  (\ref{inviscid_model}): for $x=O(\eps)$,  $y_d-y_0 =O(1)$ otherwise $y_s-y_0=O(\eps^2)$ - see Figure \ref{model}. The key point here is that the order of the correction does {\em not} change at the boundary in contrast to the Neumann case where the correction grows to $O(\eps)$.  In fact, in this model problem the correction goes to precisely zero there because the $\eps=0$ problem has a well defined boundary value. Hence, again, diffusion-free conditions are better `do least harm' conditions than Neumann.

The problem (\ref{viscous_model}) \& (\ref{d_free_bcs}) can also be solved numerically using a standard Chebyshev expansion and shows no issues - see Figure \ref{model}(right).  It's worth remarking that the same solution is obtained (up to discretization errors) if the  boundary conditions $y''(\pm 1)=0$ are used instead. These are equivalent for a smooth solution of course.

%
%

\section{\label{cylinder}Inertial wave problem in a cylinder}

We now switch our attention to a more serious PDE example. The linearised equations for describing small motions of a viscous,
incompressible fluid away from uniform rotation are
\beq
\frac{\partial \bu}{\partial t}+2 \bhz \times \bu +\bnab p=E \nabla^2\bu,    \label{1}
\eeq
\beq
\bnab . \bu=0, \label{2}
\eeq
in the rotating frame.  This is a core problem in rotating flows with appropriately more  complicated analogues when further physics is present (e.g. stratification and/or magnetic fields). It also displays a number of fascinating viscous features typified by boundary layer breakdowns which give rise to internal shear layers and a host of different viscous  lengthscales (e.g. see figure 4 of \cite{He22}).  The basic rotation rate $\Omega$ and
cylindrical radius $S$ have been used to non-dimensionalise the
system, with the Ekman number $E=\nu/\Omega S^2$ appearing as the
non-dimensionalization of the kinematic viscosity $\nu$. 

The inviscid limit, in
which $E$ is set to zero and only a no-normal velocity boundary condition needs be applied,  gives rise to the well known inertial
wave problem, where wave solutions
\[
[\bu,p]=[u(s,z),v(s,z),w(s,z),\Phi(s,z)] e^{i(m \phi+\lam t)},
\]
may be sought (e.g. see \cite{G68,Z17}), and $(s,\phi,z)$ are cylindrical coordinates). The reduction of the problem to one for only the pressure
realises the Poincar\'{e} (1910) equation
\beq
\frac{1}{s} \frac{\partial}{\partial s} s \frac{\partial \Phi}{\partial s}
- \frac{m^2}{s^2}
\Phi+
\left(1-\frac{4}{\lam^2} \right) \frac{\partial^2 \Phi}{\partial z^2}=0
\label{poincare}
\eeq
to be solved subject to the boundary conditions
\beqa
\frac{\partial \Phi}{\partial s}+\frac{2 m}{\lam s} \Phi &=& 0 \quad {\rm on}
\, s=1, \nonumber \\
\frac{\partial \Phi}{\partial z} &=& 0 \quad {\rm on} \, 
z=0,\, d. \nonumber
\eeqa
For the cylinder problem, Kelvin (1880) first wrote down the separable solutions 
\beq
\bu_{mnl}=\frac{e^{i(m \phi +\lam t)}}{2(4-\lam^2)}
\left[ \begin{array}{c}
i \{(\lam+2) J_{m-1}(ks)-(\lam-2) J_{m+1}(ks) \} \cos( l \pi z/d) \\
- \{(\lam+2) J_{m-1}(ks)+(\lam-2) J_{m+1}(ks) \} \cos( l \pi z/d) \\
2 i \lam k d (\pi l)^{-1} J_m(ks) \sin( l \pi z/d)
\end{array} \right],  \label{u}
\eeq
\beq
p=-\frac{1}{k} J_m(ks) \cos( l \pi z/d) e^{i(m \phi+\lam t)},
\label{p}
\eeq
where 
\[
\lam=\frac{\pm 2}{\sqrt{1+k^2 d^2/(\pi l)^2}},
\]
and $k$ is a solution, indexed by $n$ such that 
$0\,<\,k_{n=1}\,<\,k_{n=2}\ldots$, of
\[
s \frac{d}{ds} J_m(ks)+\frac{2 m}{\lam} J_m(ks)=0|_{s=1}.
\]
Three numbers, $l$, $n \in {\mathbb N}$, and $m \in {\mathbb Z}$, are sufficient to
specify the inertial mode and in particular the frequency
$\lam=\lam_{lmn}$.  These indices correspond roughly with the number
of nodes axially, radially, and azimuthally respectively in the
pressure eigenfunction. As in KB95 and \cite{K99},  
we focus here on the inertial mode with $m=n=l=1$ and $d=1.9898$ (the quoted value in KB95) with inertial velocity field $\bu_0$ and real frequency $\lambda_0 =1.000007648337 \approx 1$. The subscript is to indicate that these are the $E=0$ quantities compared to those, $\bu_E$ and $\lambda_E$, for $E \neq 0$. The viscous correction is then defined as $\bu_c:=\bu_E-\bu_0$. Reinstating viscosity requires 2 extra (tangential) boundary conditions to supplement the no-penetration (normal) condition.

\subsection{Non-slip boundary conditions}

If the container walls are rigid then non-slip boundary conditions are required which is the usual case discussed in text books (e.g.  \S 2.5 in \cite{G68}).  If $\bu_0(\bx)e^{i \lambda t}$ is the solution to the inertial wave problem (\ref{poincare}) with $\lambda$ a real frequency in the interval $[-2,2]$ for $E=0$,   then the leading viscous correction for $E \neq 0$ is the $O(E^0)$ boundary velocity $\btu_0(\bx,t)$  forced by the non-slip condition
\beq
\bu_0+\btu_0={\bf 0}|_{\partial V}.
\eeq
(hereafter $\mathbf{\tilde{\,\,}}$ indicates a boundary layer flow).
This decays into the interior from the boundary over an Ekman boundary of thickness $E^{1/2}$ but drives an $O(E^{1/2})$ viscous flow $\bu_1(\bx,t)$ in  the interior by `Ekman pumping' (equation 2.6.13 in \cite{G68}).  These flows mean that $\lambda$ becomes complex with a viscous decay rate $\lambda_i:=\Im m(\lambda)=O(E^{1/2}) \,>\,0$.  There are also  shear layers spawned by the corners of the cylinder (e.g. \cite{McEwan70} and figures \ref{E4}-\ref{E6} below) but these appear unimportant for the leading decay rate calculation (\S2.9 in \cite{G68}).

\subsection{Stress-free boundary conditions}

For other situations,  the boundary may be stress-free (e.g. for a gaseous planet) or this choice is assumed to make the viscous correction at the boundary as weak as possible (e.g. \cite{Lin21, Vidal23}). Then the leading viscous correction is $E^{1/2}\btu_1(\bx,t)$ in a boundary layer of thickness  $E^{1/2}$ such that, at, for example, the cylinder sidewall,
\beq
s\frac{\partial}{\partial s} \biggl( \frac{v_0}{s} \biggr)-\frac{\partial \tilde{v}_1}{\partial \xi}=0 \biggl|_{s=1}   \biggr.
\quad \& \quad
\frac{\partial w_0}{\partial s}-\frac{\partial \tilde{w}_1}{\partial \xi} = 0 \biggl|_{s=1}  \biggr.
\eeq
where $\xi:=(1-s)/E^{1/2}$ is the rescaled normal coordinate.  The corrective viscous {\em bulk} flow (as opposed to localised shear layers) is then $O(E)$ driven by the subsequent $O(E)$ Ekman pumping at the boundaries and the bulk viscous term also of $O(E)$.
As a result the viscous decay rate is much smaller - $\lambda_i:=O(E) \,>\,0$.

Unfortunately, for our purposes here, the cylinder has a singular feature for these boundary conditions: the inertial mode automatically satisfies stress-free conditions on the top and bottom boundaries, i.e. $\partial u_0/\partial z=\partial v_0/\partial z=0$ at $z=0$ and $z=d$  already.  Hence a viscous boundary layer only forms on the side wall and in particular no internal shear layers are formed by the corner region where the horizontal boundaries meet the sidewall which is an exceptional response. - e.g.  a calculation in more geophysically-interesting sphere or spherical shell  does have shear layers (e.g. \cite{Lin21,Vidal23} and the next section). As a result,  for stress-free boundary conditions, the calculation for the viscous decay rate  $\lambda_i$ in a cylinder is just 1D - the variation in $z$ can be separated off - and so  very small Ekman numbers can be reached easily (see Table \ref{decayrates}).

\subsection{Diffusion-free boundary conditions}

Diffusion-free boundary conditions are 
\beq
i \lam u-2v+\frac{\partial p}{\partial s}  =0, \quad 
i \lam v+2u+\frac{imp}{s} =0 \quad \& \quad
w =0
\label{top/bottom}
\eeq
on the top ($z=d$) and bottom ($z=0$)
and
\beq
u = 0, \quad
i \lam v+2u+\frac{imp}{s} =0 \quad \& \quad
i \lam w +\frac{\partial p}{\partial z}=0 \label{sides}
\eeq
 on the sides ($s=1$) of the cylinder. For a $C^2$ solution this is equivalent to setting the tangential components of the  vectorial Laplacian to zero. That is, the conditions
\beq
u =0, \nonumber 
\eeq
\beq
 \bhphi.\nabla ^2\bu =\frac{\partial^2 v}{\partial s^2}+\frac{1}{s} \frac{\partial v}{\partial s}-\frac{m^2+1}{s^2}v +\frac{\partial^2 v}{\partial z^2} +\frac{2 i m u}{s^2} =0, \nonumber 
\eeq
\beq
\bhz    .\nabla^2 \bu =\frac{\partial ^2w}{\partial s^2}+\frac{1}{s} \frac{\partial w}{\partial s}-\frac{m^2}{s^2} w +\frac{\partial^2 w}{\partial z^2} =0
\label{sides_again}
\eeq
imposed at $s=1$ are equivalent to those in (\ref{sides}) having the same complex frequencies $\lambda(E)$ as eigenvalues (not shown but verified).

\subsection{Scalings}

The inviscid inertial wave satisfies all the diffusion-free boundary conditions. However the $O(E)$ bulk viscous term drives a primary $O(E)$  interior flow, $E \bu_2(\bx,t)$, which does not, leading to an  $O(E)$ error in the diffusion-free boundary conditions (with no boundary-normal derivatives). These can be fixed up by a very weak $O(E)$ boundary correction flow, $E \btu_2(\bx,t)$. The associated Ekman pumping drives a secondary  bulk  flow at $ O(E^{3/2})$. Hence the corrective viscous flow   is $O(E)$ in {\em both} the bulk and the Ekman boundary layer. In contrast,  the secondary flow is $O(E)$ in the bulk and a larger $O(E^{1/2})$ in the boundary layer for stress-free boundary conditions.  For non-slip conditions, the secondary flow is $O(E^{1/2})$ in the bulk and $O(1)$ in the boundary layer.


%
%
%
\subsection{Numerical Implementation}

We now demonstrate the scaling predictions discussed above using numerical calculations over a range of small but finite $E$. As in KB95,  we exploit the symmetries of the inertial waves by only solving the eigenvalue problem (\ref{1})-(\ref{2}) in the upper `quadrant' $(s,z) \in [0,1] \times [d/2,d]$ of an extended cylinder $(s,z) \in [-1,1] \times [0,d]$.  If $m$ (the azimuthal wavenumber) is even (odd),  then $(w, p)$ are even (odd) and $(u,v)$ are odd (even) functions of $s$. Likewise about the cylindrical equator $z = d / 2$,  if $l$ is even (odd),  then $(u, v,p)$ are even (odd) and $w$ is  odd (even) functions of
$z^* := 2z/d- 1$. Hence,  for the situation of interest here,  $m = l = 1$, the appropriate spectral expansions used are
\beq
(u,v) (s,z) = \sum^N_{i=1} \sum_{j=1}^N ( u_{ij}, v_{ij} ) T_{2i-2}(s) T_{2j-1}(z^*), 
\eeq
\beq
w(s,z)=\sum^N_{i=1} \sum_{j=1}^N w_{ij} T_{2i-1}(s)T_{2j-2}(z^*), \qquad p(s,z)= \sum^N_{i=0} \sum_{j=0}^N p_{ij} T_{2i-1}(s)T_{2j-1}(z^*)
\eeq
where $T_n(x):=\cos( n \cos^{-1}x)$ is the $n^{th}$ Chebyshev polynomial. The equations were collocated over the positive zeros of $T_{2N}$ in $s$ and $z^*$ (the expansion lengths in $r$ and $z$ are taken equal for simplicity).  Inverse iteration was then used to find the relevant (complex) eigenvalue and eigenfunction as the inviscid real frequency $\lambda_0$ provides a good guess.  KB95 talk about a maximal  (given the 300 MBytes available) $N=35$ full eigenvalue solve taking 3 days on a Sun SparcCenter 2000 (the University of Newcastle's largest computer in 1995).  The equivalent today takes 9.4 mins on an Apple M1 laptop even using Matlab instead of the Fortran 77 of KB95.  Employing inverse iteration to just focus on the eigenvalue of interest brings that down to just 18 seconds.

%
%
\begin{table}
\caption{\label{decayrates}  Decay rates $\lam_i$ at various $E$ and different boundary conditions. Non-slip and diffusion-free boundary conditions use up to $N=110$ to check convergence. The stress-free case reduces to a 1D problem and resolution can go much higher:  $N=6,000$ and $10,000$ give $\lambda= 1.0000076483366+i9.469658E$ for $E=10^{-10}$.  Removing the top and bottom boundary conditions for the diffusion-free case gives the starred decay rates on the far right which shows  convergence to the interior dissipation rate value of $9.97105E$ indicating that the boundary layer contribution is higher order.}
\begin{center}
\begin{tabular}{@{}lrlccrccll@{}}
&&&&&&&&&\\
\hline\noalign{\smallskip}
&&&&&&&&&\\
& $E$                     & \hspace{0.35cm}  &   KB95        & \hspace{0cm}     &    non-slip                          &\hspace{1cm} &       stress-free        & \hspace{0cm} &     diffusion-free  \\
&                            &                               &                     &   &                                    &                            &                  &                              &          \\
&                              &                            &  $\lam_i/\sqrt{E}$  & &    $\lam_i/\sqrt{E}$           &                           &     $\lam_i/E$    &                             &  $\lam_i/E$ \\
&                             &                             &                         & &                                     &                           &                   &                            &      \\\hline
&&&&&&&&&\\
& $10^{-4}$           &                             &  $ 1.5665$    & &    $ 1.5665$   &                           &     $9.4362$ &                     &    $9.831$\quad $ $  \\
& $10^{-5}$              &                          &  $ 1.4834$    &  &   $ 1.4859$   &                           &     $9.4591$ &                      &    $9.927$ \quad $ $ \\
& $10^{-6}$               &                         &                        &   &  $ 1.4619$    &                           &    $9.4663$ &                      &   $9.957$  \quad $9.9647* $\\
& $3 \times 10^{-7}$ &                         &                        &  &   $1.4571$    &                           &                      &                      &   $9.963$   \\
& $10^{-7}$              &                           &                       &  &                       &                            &   $9.4686$ &                  &   \hspace{1cm}\quad $9. 9689^*$ \\
& $10^{-8}$              &                           &                       &  &                       &                            &   $9.4694$ &                  &  \hspace{1cm}\quad $9.9703^* $  \\
& $10^{-9}$              &                           &                       &  &                       &                            &   $9.4696$ &                  &  \hspace{1cm}\quad $9.9707^* $  \\
& $10^{-10}$             &                           &                       &  &                       &                            &   $9.4697$ &                  &   \hspace{1cm}\quad $9.9708^* $ \\
& $10^{-11}$             &                           &                       &  &                       &                            &                     &        &   \hspace{1cm}\quad $9.9709^* $ \\
&                                &                           &  \multicolumn{3}{c}{$ \underbrace{\hspace{3.25cm}}_{E\,\rightarrow\, 0}$}  &                
     &        \multicolumn{3}{c}{$ \underbrace{\hspace{4.75cm}}_{\text{interior dissipation only}}$}                   \\
&                                 &                          &                        &   &                                                       &                        &                     &                     &              \\    
&Wedemeyer            &                         &  \multicolumn{3}{c}{$1.732$}       &                           &           \multicolumn{3}{c}{$9.97089724$}               \\ 
& Kudlick                   &                         &  \multicolumn{3}{c}{$1.451$}       &                           &                 &                           & \\
&&&&&&&&&\\
\noalign{\smallskip}\hline
\end{tabular}
\end{center}

\end{table}

%
%
\begin{table}
\caption{\label{Convergence}  Convergence of decay rate $\lam_i/\sqrt{E}$ estimates with varying $N$ for non-slip boundary conditions ($N=110$ is the largest resolution that runs  on a  192GB memory machine).}
\begin{center}
\begin{tabular}{@{}lrlccrccll@{}}
&&&&&&&&&\\
\hline\noalign{\smallskip}
&&&&&&&&&\\
& $N$   & \hspace{0.5cm} &   $E=10^{-4}$ &\hspace{0.5cm} &    $E=10^{-5}$  &\hspace{0.5cm} &    $E=10^{-6}$    & \hspace{0.5cm} &     $E=3\times 10^{-7}$  \\
&           &                             &                            &   &                            &             &                  &                              &          \\ \hline
&&&&&&&&&\\
&   30   &                             &  $ 1.5664732$ & &    $1.4861012$ &              &    &                           &      \\
&   40   &                             &  $ 1.5665071$ &  &   $1.4870688$&              &                             &   &       \\
&   50   &                             &  $ 1.5665072$ &  &   $1.4866554$&              &   $1.4396532$  &   &      \\
&   60   &                             &   $1.5665072$ &  &   $1.4865893$&              &   $1.4684427$ &   &  $1.3328341$    \\
&   70    &                            &    $1.5665072$ &   &   $1.4865937$&               &   $1.4623715$  &   &   $1.4502321$  \\
&   80    &                            &                           &   &    $1.4865936$&               &   $1.4622908$  &   &   $1.4639174$   \\
&   90    &                            &                           &  &      $1.4865936$&               &   $1.4619161$   &   &   $1.4581648$  \\
&  100    &                           &                           &   &                            &                &   $1.4618646$  &   &   $1.4571169$  \\
 & 110    &                           &                           &   &                             &                &   $1.4618879$  &   &   $1.4571179$  \\
 &&&&&&&&&\\
 \noalign{\smallskip} \hline
\end{tabular}
\end{center}
\end{table}

%
%
\begin{figure}
\centering
\scalebox{0.5}[0.5]{\includegraphics{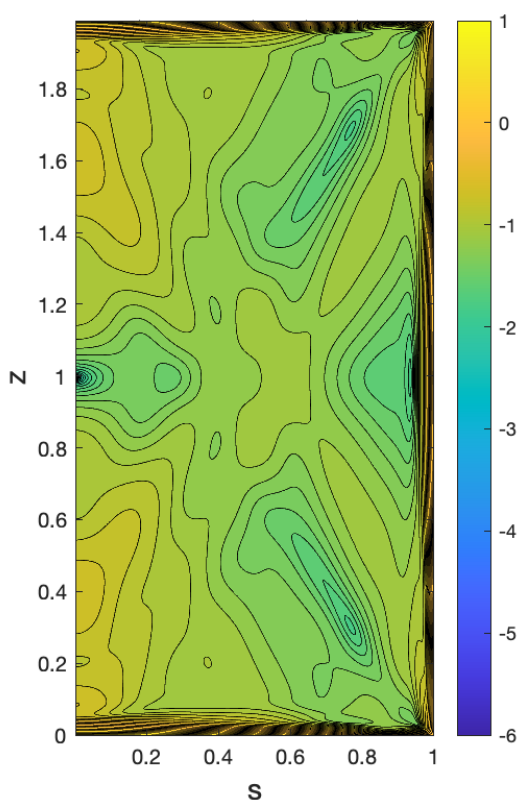}}
\scalebox{0.5}[0.5]{\includegraphics{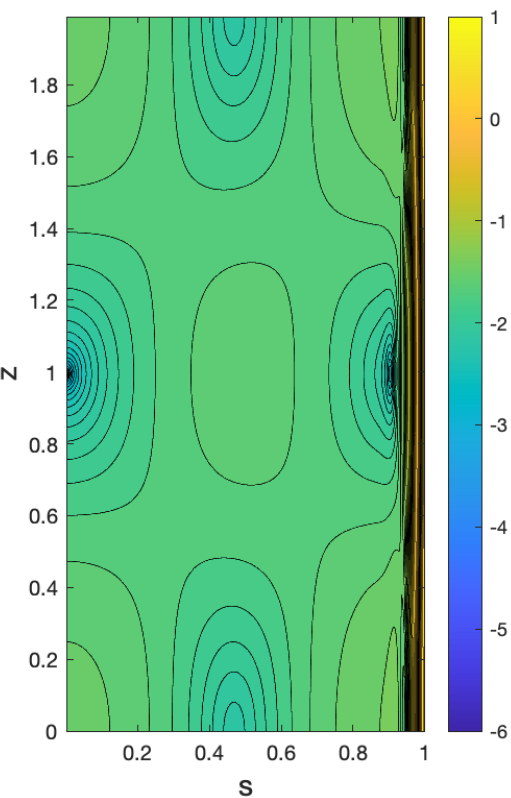}}
\scalebox{0.5}[0.5]{\includegraphics{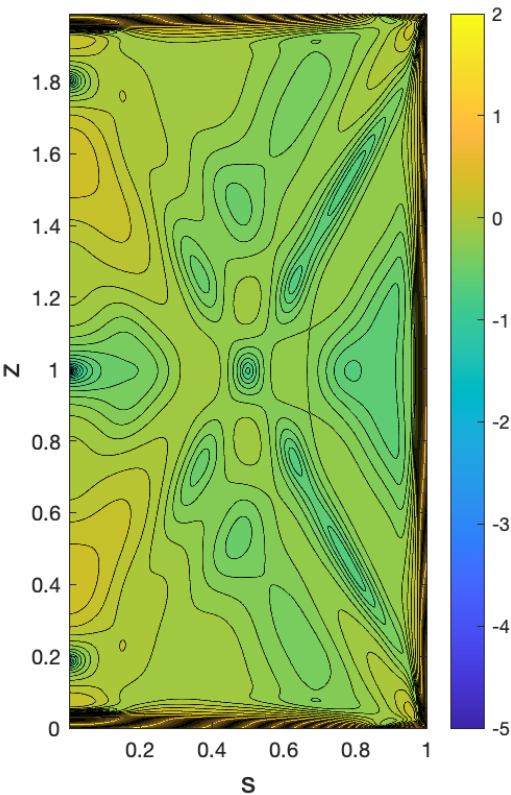}}
\caption{\label{E4} Plots of the viscous corrective field for different boundary conditions at $E=10^{-4}$: non-slip (left) $\log_{10} |\bu_c| $ plotted with $\max |\bu_c|=4.24$,  stress-fress (middle) $\log_{10} |\bu_c|/\sqrt{E}$ plotted with  $\max |\bu_c| = 3.03\sqrt{E}$ and diffusion-free conditions (right) $\log_{10} |\bu_c|/E$ with $\max |\bu_c| = 44.56E$. This and figures \ref{fig3} and \ref{fig4}  show how the viscous corrective flows scale with $E$ for the various boundary conditions.
}
\label{fig2}
\end{figure}

%
%
\subsection{Decay rate $\lambda_i$ and Eigenfunctions}

%
%
To identify the viscous corrective flow to the inviscid inertial wave $\bu_{111}$ - see (\ref{u}) - due to $E \neq 0$, we use the well-known orthogonality property of inertial waves (\S2.7  \cite{G68}) to renormalise the viscous eigenfunction to have exactly the inertial wave field $\bu_{111}$ and then subtract it - 
\beq
\bu_c:= \biggl[\frac{\int_V \bu^*_{111}. \bu_{111} \, {\bf d}^3 \bx}{\int_V \bu^*_{111}. \bu_{E} \, {\bf d}^3 \bx} \biggr] \bu_E-\bu_{111}.
\label{renormalisation}
\eeq
This definition means that $\bu_c$ is orthogonal to $\bu_{111}$ and is  normalised to be the viscous correction to the field $\bu_{111}$ as defined in (\ref{u}) 

Table \ref{decayrates} collects all the decay rate results across the 3 different boundary conditions mentioned above. Only the non-slip case was considered in KB95 and the results there are confirmed for $E=10^{-4}$ and  now properly converged for $10^{-5}$. Treating even smaller $E$ suggests that Kudlick's formula for the decay rate is to be preferred to Wedemeyer's: see KB95 for context.  Table \ref{Convergence} shows how the non-slip results converge. The $N=30$ result for $E=10^{-5}$ is misleadingly good compared to KB95.  In fact $N=35$ gives $\lambda_i=1.4913$ before convergence really starts at $N=40$.
%
%
\begin{figure}
\centering
\scalebox{0.5}[0.5]{\includegraphics{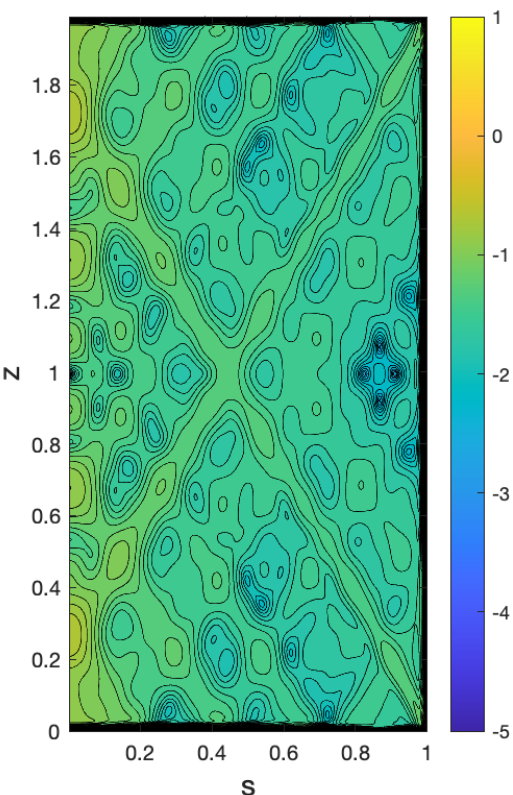}}
\scalebox{0.5}[0.5]{\includegraphics{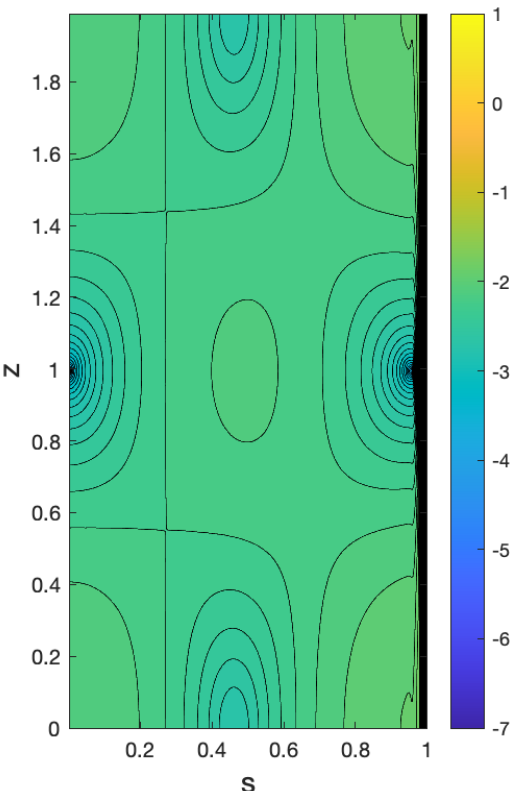}}
\scalebox{0.5}[0.5]{\includegraphics{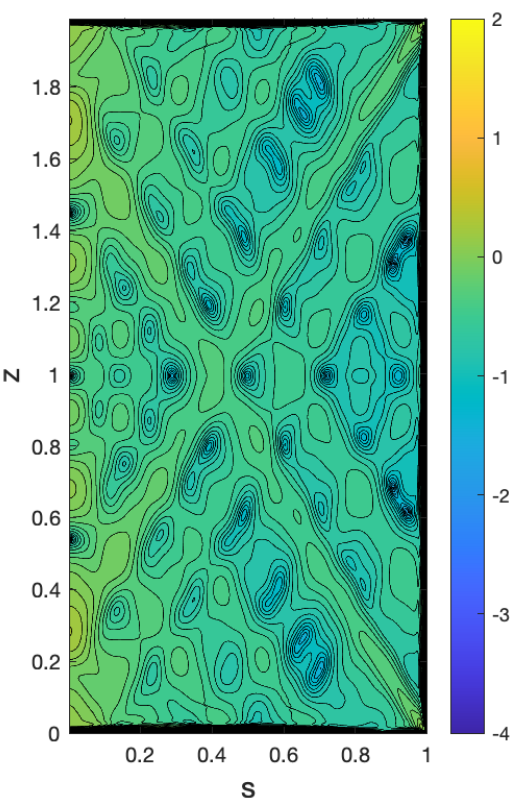}}
\caption{\label{E5} Plots of the viscous corrective field for different boundary conditions at $E=10^{-5}$: non-slip (left) $\log_{10} |\bu_c| $ plotted with $\max |\bu_c|=4.24$,  stress-fress (middle) $\log_{10} |\bu_c|/\sqrt{E}$ plotted with  $\max |\bu_c| = 3.00\sqrt{E}$ and diffusion-free conditions (right) $\log_{10} |\bu_c|/E$ with $\max |\bu_c| = 43.47E$. This and figures \ref{fig2} and \ref{fig4} show how the viscous corrective flows scale with $E$ for the various boundary conditions.
}
\label{fig3}
\end{figure}

%
%
\begin{figure}
\centering
\scalebox{0.5}[0.5]{\includegraphics{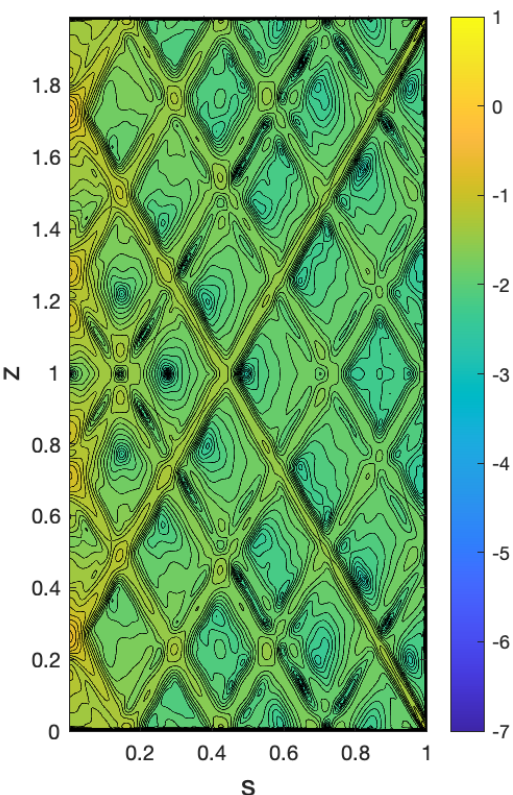}}
\scalebox{0.5}[0.5]{\includegraphics{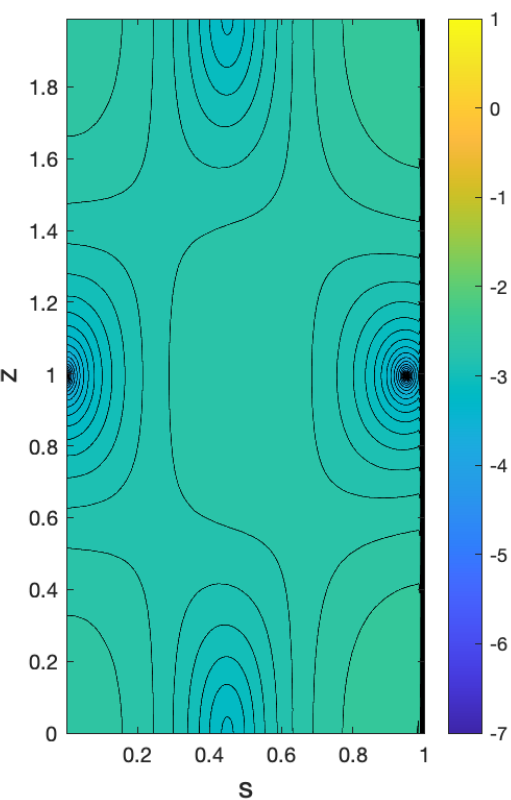}}
\scalebox{0.5}[0.5]{\includegraphics{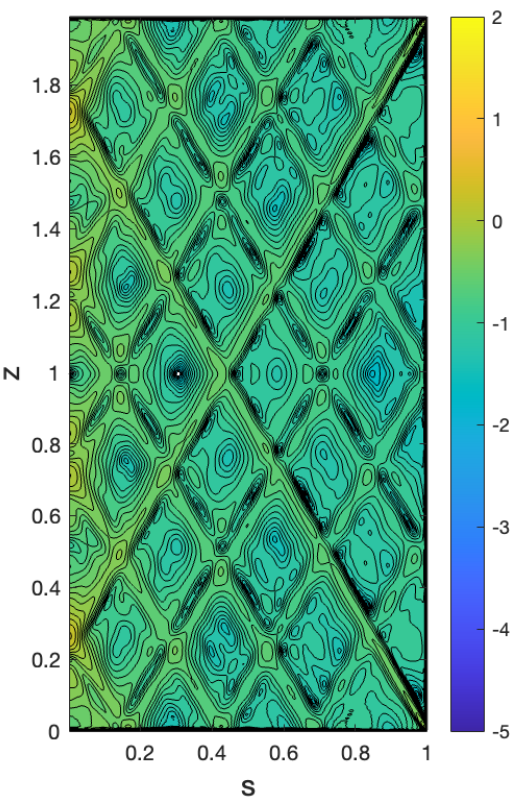}}
\caption{\label{E6} Plots of the viscous corrective field for different boundary conditions at $E=10^{-6}$: non-slip (left) $\log_{10} |\bu_c| $ plotted with $\max |\bu_c|=4.24$,  stress-fress (middle) $\log_{10} |\bu_c|/\sqrt{E}$ plotted with  $\max |\bu_c| =2.99\sqrt{E}$ and diffusion-free conditions (right) $\log_{10} |\bu_c|/E$ with $\max |\bu_c| = 42.35E$. This and figures \ref{fig2} and \ref{fig3} show how the viscous corrective flows scale with $E$ for the various boundary conditions.
}
\label{fig4}
\end{figure}

The contribution to the decay rate from the interior can be estimated as $(k^2+\alpha^2)E$ (\cite{K99}) and with $\lambda^2=4/(1+k^2/\alpha^2)$ is just $4\alpha^2 E/\lambda^2=9.97089724E$.
The diffusion-free decay rate is well approximated by
\beq
\lam_i = 9.97089724E-14E^{3/2}+O(E^2)
\eeq
indicating that the boundary layer contribution is $O(E^{1/2})$ smaller than the interior's (this is clearer for the 1D case where the top and bottom boundaries are made stress-free - see the starred entries in Table 1 which can be computed to much smaller $E$).  This is in contrast to the stress-free case where
the decay rate does not converge to the interior's contribution as $E \rightarrow 0$ - see Table 1 - indicating that the boundary layer contribution is at the same order as the interior's.  Hence the diffusion-free boundary condition again is less disruptive than the stress-free (Neumann) condition.

%
%
\begin{table}
\caption{\label{maxvel} Maximum amplitude of the viscous correction velocity $|\bu_c(r,z)|$ over the cylindrical domain. }
\begin{center}
\begin{tabular}{@{}lrlccrcc@{}}
&&& &&& &\\
\hline\noalign{\smallskip}
&&& &&& &\\
& $E$  & \hspace{0.5cm}  &  non-slip  & \hspace{0.25cm} &   stress-free   &  \hspace{0.25cm} &   diffusion-free  \\
&&&  &&& & \\ 
\hline\noalign{\smallskip}
&&& &&& &\\
&   $10^{-4}$  & &   $4.24$      &  &  $3.03\sqrt{E}$ & &  $44.56E$ \\
&   $10^{-5}$  & &   $4.24$      &  &  $3.00\sqrt{E}$ & &  $43.48E$ \\
&   $10^{-6}$  & &   $4.24$      &  &  $2.99\sqrt{E}$ & &   $42.35E$ \\
&&& &&& &\\
\noalign{\smallskip}\hline
\end{tabular}
\end{center}
\end{table}
%
%
\begin{figure}
\centering
\scalebox{0.6}[0.6]{\includegraphics{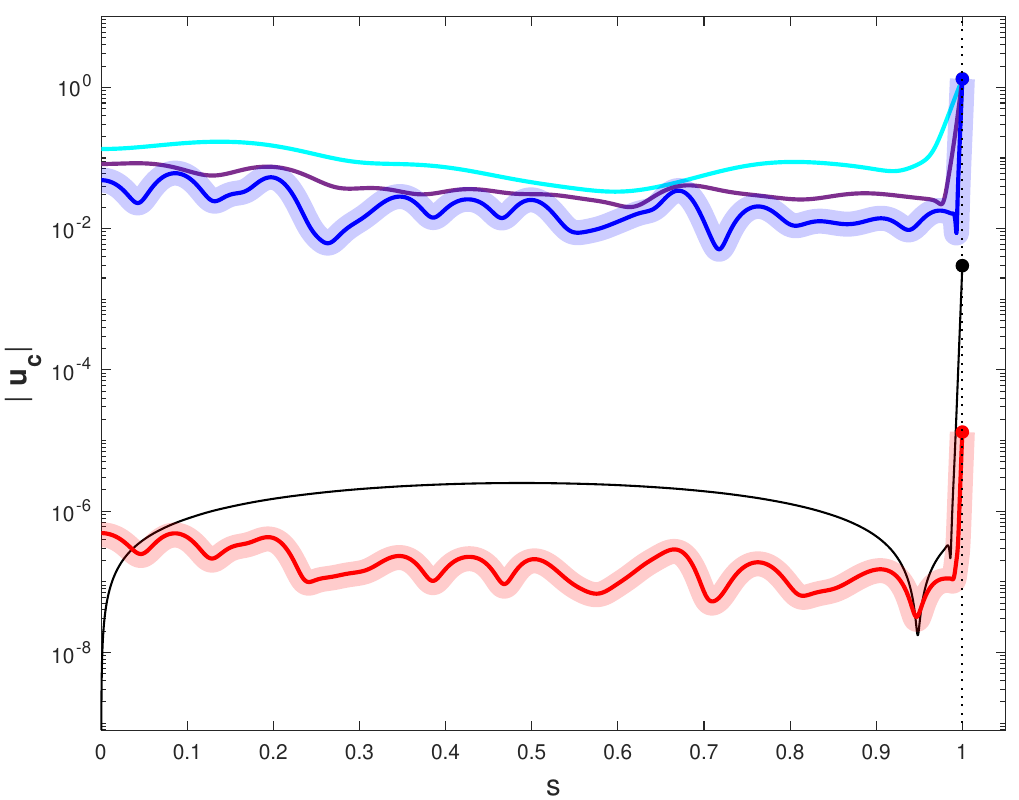}}
\caption{\label{lineplots}  The magnitude of the {\em unscaled} corrective viscous velocity, $|\bu_c|(s,z^*)$,  plotted as a function of  $s$ for a fixed choice of $z^*$ and  various different boundary conditions.  The upper 3  (blue,  purple and cyan) lines are for non-slip conditions with $z^*=d/\sqrt{2}$ (chosen to be arbitrary). The wide transparent  blue line corresponds to $N=80$ and the blue line $N=90$ both at $E=10^{-6}$ confirming that the resolution is sufficient; the purple line is for $E=10^{-5}$  and the cyan line is for $E=10^{-4}$ (the corresponding dots at $s=1$ are all at $1.32$).   The middle black line is for stress-free conditions at $E=10^{-6}$ ($N=100$) and uses $z^*=d/2$ as the interior happens to rather featureless at $z^*=d/\sqrt{2}\approx 1.4$ - e.g.  see figure \ref{E6}(middle) (the black dot is at $2.99 \times 10^{-3}$). The lower red curves are for diffusion-free conditions with $z^*=d/\sqrt{2}$ again.  The wide transparent red line uses $N=90$ and the thin red line threading it has $N=110$ (the red dot is at $1.32\times 10^{-5}$).  This plot emphasizes the difference in scales of the boundary-induced flows between the different boundary conditions and how the interior structure starts to emerge at $E=10^{-6}$ - note the structure for the diffusion-free conditions is very similar to that for the non-slip conditions albeit 4 orders or magnitude smaller.  }
\label{fig6}
\end{figure}

The various corrective viscous velocities $\bu_c(r,z) $ are shown in Figures \ref{E4}-\ref{E6} for $E=10^{-4}$, $10^{-5}$ and $10^{-6}$ for the different  choices of boundary conditions. Due to the vastly different amplitudes of these flows, the leading viscous scaling with $E$ is removed before contouring. That is, for non-slip boundary conditions, we expect  $\bu_c$ to be $O(1)$ in the boundary layer so there is no rescaling in this case whereas for the stress-free case, $\bu_c$ is $O(E^{1/2})$, so $\log_{10} |\bu_c/E^{1/2}|$ is plotted and  $\log_{10} |\bu_c/E|$ for the diffusion-free case. The various maximum amplitudes are collected in Table \ref{maxvel} confirming the expected scaling. The formation of internal shear layers is already evident at $E=10^{-4}$ for the non-slip and diffusion-free conditions and these get more intense and narrower as $E$ decreases (their thickness is $E^{1/3}$).  Unfortunately for our purposes here, they are completely absent from the stress-free situation for reasons already mentioned.

Figure \ref{lineplots} shows  the {\em  unscaled} amplitude of $\bu_c$ as a function of $s$ at a given $z$.  The top three curves are for the non-slip situation for $E=10^{-4}$, $10^{-5}$ and $10^{-6}$ and show how the internal shear layers only really become prominent at $E=10^{-6}$. All have the same $O(1)$ value at the side wall ($=-|\bu_{111}(1,z)|$).  The thin black line is the  stress-free profile at $E=10^{-6}$ which has an $O(E^{1/2})$ boundary value and is largely featureless in the interior as already commented on.  Finally the lowest thick red line (computed with $N=90$) and the thin red line through its centre (computed with $N=110$) are for the diffusion-free conditions.  This is an $O(E^{1/2})$ smaller than the non-slip flow {\em everywhere}.  Perhaps, not surprisingly, the velocity profiles are largely similar in shape as they are formed by the same geometrical effects (the reflections of the internal shear layers) but the point is, of course, their magnitude is very different.

Finally, we note that higher derivatives than second order can be set to zero at the boundaries to weaken the boundary layers even though ordinary (second order) diffusion is being used: see Appendix B. The philosophy of diffusion-free boundary conditions can be readily applied with hyperviscosity  as well. In this case, further conditions are also needed and these can be systematically defined as integrals or differentials of the diffusion-free b.c.s.: again see Appendix B.

%
%
%

%
%

%
%

%
%
\section{Inertial wave problem in a sphere}
%

The inertial wave problem in a sphere (equations (\ref{1}) and (\ref{2})) also has separable solutions in the inviscid limit where only the normal velocity needs to vanish at the walls. These separable solutions involve modified spheroidal coordinates ($\eta, \mu$) introduced by \cite{Bryan1889}
where
\beq
    s=\left(\frac{4}{4-\lambda^2}-\eta^2\right)^{1/2}(1-\mu^2)^{1/2}, \quad    z=\left(\frac{4}{\lambda^2}-1\right)^{1/2}\eta \mu.
\eeq
Using these, the pressure field can be written as 
(e.g. see \cite{G68,Z17})
\begin{equation}
\Phi_{mnk}=P_n^m(\eta/c_{mnk})P_n^m(\mu)\mathrm{e}^{\mathrm{i}(m\phi+\lambda t)}
\end{equation}
where $P_n^m$ represents an associated Legendre function and  $c_{mnk}=(1-\lambda_{mnk}^2/4)^{-1/2}$. The real frequency $\lambda_{mnk}$ is the $k$th eigenvalue solution of the transcendental equation
\beq
mP_n^m(\lambda/2)=2\left(1-\frac{\lambda^2}{4}\right)\frac{d}{d\lambda}P^m_n(\lambda/2),
\eeq
and so each inertial mode is uniquely specified by a triplet of indices $(m,n,k)$.\\

\subsection{Numerical Implementation}

The viscous problem ($E \neq 0$) is numerically solved using a pseudo-spectral method based on an expansion of spherical harmonics on spherical surfaces and Chebyshev collocation in the radial direction.
The incompressible condition $\nabla \cdot \bu =0$ is fulfilled by the the toroidal-poloidal decomposition 
\beq
\bu=\bnabla \times (\TT \bR)+\bnabla\times \bnabla \times (\PP \bR) \label{eq:TP}
\eeq
and, for an inertial wave with the azimuthal wavenumber $m$, the scalars $\TT$ and $\PP$ are expanded in spherical harmonics
\beq
\TT=\sum_{n=m}^N r\TT_n(r)Y_n^m(\theta,\phi), \qquad
\PP=\sum_{n=m}^N r^2 \PP_n(r)Y_n^m(\theta,\phi). \label{eq:PP}
\eeq
where $(r,\theta,\phi)$ are spherical polar coordinates. The additional factors of $r$ and $r^2$ in the above expansions are used to match the regularity conditions at the origin of a full sphere.
The projections $\bR \cdot \bnabla \times$ and $\bR \cdot \bnabla \times \bnabla \times$ of the equation 
\beq
\mathrm{i} \lambda \bu + 2\bhz \times \bu + \bnab p = E \nabla^2 \bu, \label{eq:NS}
\eeq
onto spherical harmonics, give a set of ODEs for $\TT_n(r)$ and $\PP_n(r)$ with a block tri-diagonal matrix structure. For the radial dependence, Chebyshev collocation is used on the $K+1$ Gauss-Lobatto nodes with the projected ODEs replaced by corresponding boundary conditions on the boundary nodes. 

\subsection{Boundary Conditions}

The non-slip boundary conditions $\bu={\bf 0}$ correspond to 
\beq
\TT_n(r)=0, \quad \PP_n(r)=0 \quad \& \quad\frac{d \PP_n(r)}{dr}=0
\eeq
at the boundary $r=1$ (the sphere radius $R$ is used to define the Ekman number as $E:=\nu/\Omega R^2$).

The stress-free boundary conditions are
\beq
u_r=0, \quad \frac{\partial }{\partial r}\left(\frac{u_\theta}{r}\right)=0, \quad \frac{\partial }{\partial r}\left(\frac{u_\phi}{r}\right)=0,
\eeq
which translates into 
\beq
\PP_n(r)=0,\quad r^2\frac{d^2\PP_n(r)}{dr^2}+2r\frac{d\PP_n(r)}{dr}=0, \quad r\frac{d\TT_n(r)}{dr}-\TT_n(r)=0,
\eeq
for the toroidal and poloidal components on $r=1$.

The diffusion-free boundary conditions are
\beq \label{eq:dfbc_phy}
\bu\cdot \bR=0 \quad \& \quad  (\bnab^2 \bu)\times \bR= {\bf 0}
\eeq
at $r=1$ or
\beqa
\PP_n(r)=0, \\
r^2\frac{d^3\PP_n(r)}{dr^3}+6r\frac{d^2\PP_n(r)}{dr^2}-(n^2+n-6)\frac{d\PP_n(r)}{dr}=0, \\
r^2\frac{d^2\TT_n(r)}{dr^2}+2r\frac{d\TT_n(r)}{dr}-n(n+1)\TT_n(r)=0.
\eeqa
At the origin $r=0$, we apply the regularity conditions \citep{Reese2006} 
\beq
\PP_n=0, \quad  \TT_n=0,
\eeq
for the even degree of spherical harmonics and
\beq
\frac{d\PP_n}{dr}=0, \quad  \frac{d\TT_n}{dr}=0,
\eeq
for the odd degree of spherical harmonics.
The generalized eigenvalue problem is solved using an iterative method to find the least damped eigenmode around the inviscid eigenfrequency. 

%
%
\begin{figure}
\centering
\scalebox{0.7}[0.7]{\includegraphics{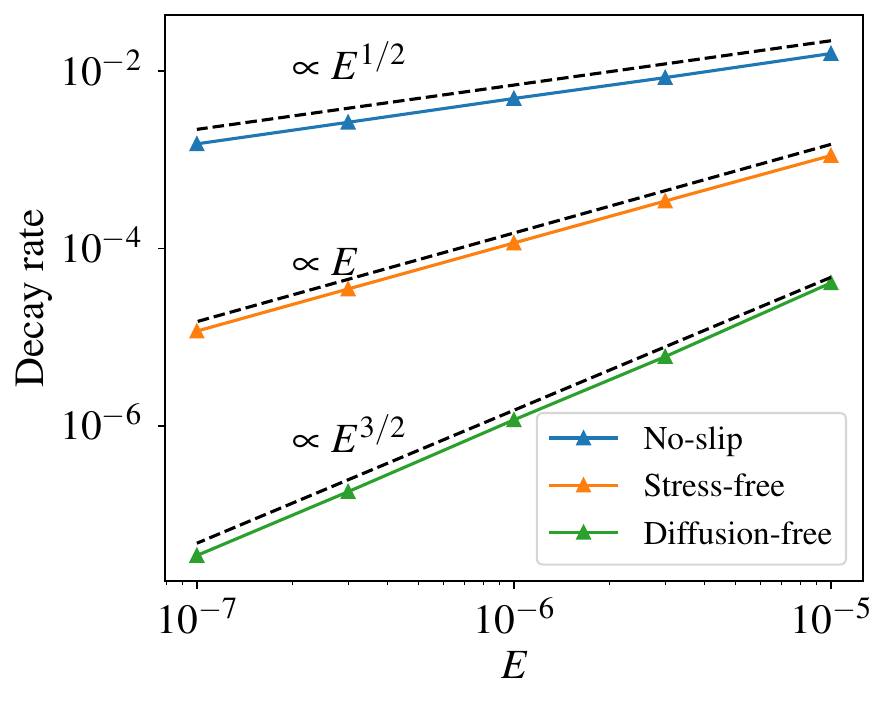}}
\caption{\label{sphere} Decay rate of inertial mode (2,4,1) with the inviscid eigen-frequency $\lambda_{241}=-1.0926$ in a sphere. }
\end{figure}
%
%
\begin{figure}
\centering
\scalebox{0.4}[0.4]{\includegraphics{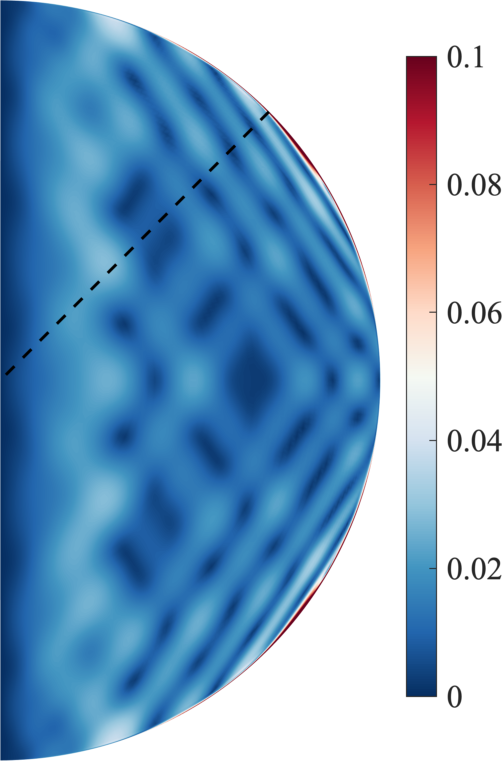}}
\scalebox{0.4}[0.4]{\includegraphics{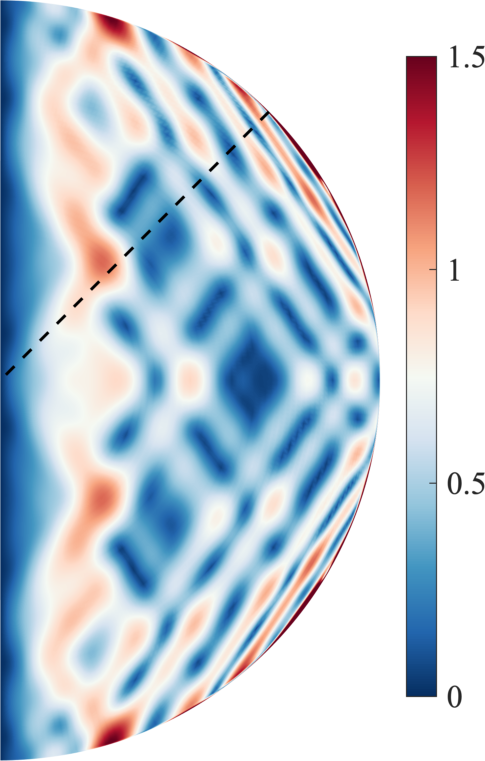}}
\scalebox{0.4}[0.4]{\includegraphics{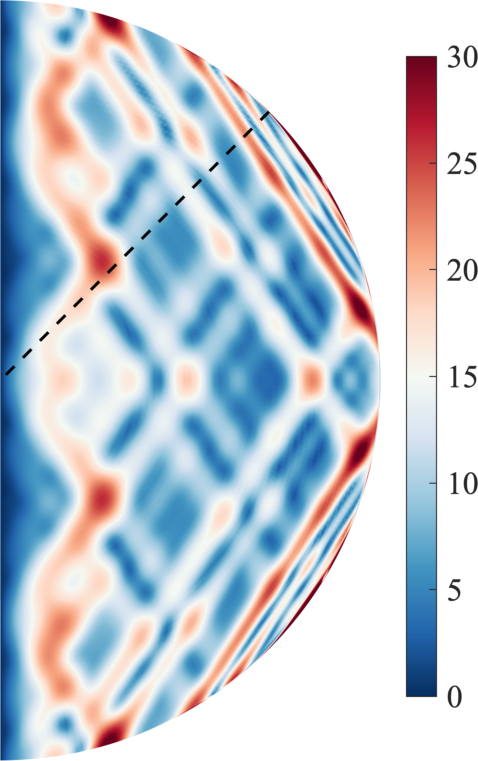}}
\caption{\label{ucsphere} Viscous corrective field of the full sphere inertial mode (2,4,1) in the meridional plane for no-slip $|{\bf u}_c|$(left), stress-free $|{\bf u}_c|/\sqrt{E}$ (middle), and diffusion-free $|{\bf u}_c|/E$ (right) boundary conditions at $E=10^{-6}$.}
\end{figure}

%
%
\begin{figure}
\centering
\scalebox{0.7}[0.7]{\includegraphics{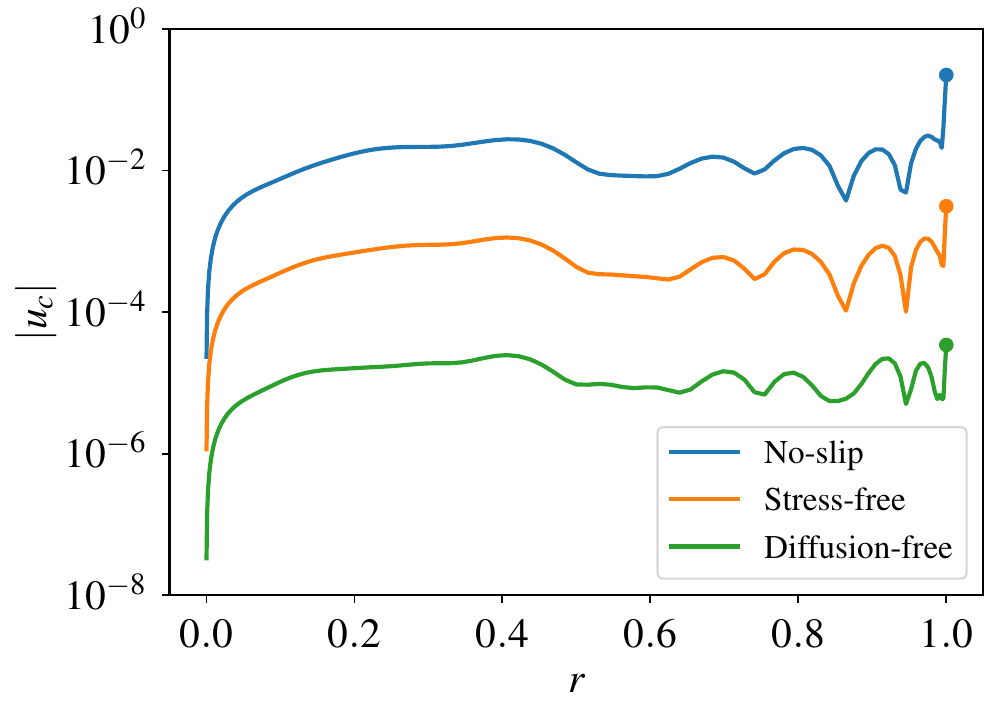}}
\caption{\label{VelCorsphere} The magnitude of the unscaled corrective viscous velocity along the dashed lines in Figure \ref{ucsphere} for a sphere at $E=10^{-6}$.}
\end{figure}

\subsection{Decay rate $\lambda_i$ for a sphere}

Figure \ref{sphere} shows the decay rate of a typical inertial mode (here $(m,n,k)=(2,4,1)$) as a function of the Ekman number for the different boundary conditions.
We find $O(E^{1/2})$ decay rates  for non-slip; $O(E)$ for stress-free and $O(E^{3/2})$ for diffusion-free b.c.s.
This indicates that the viscous correction in the diffusion-free situation is weakest.

Since there are $O(E)$ corrective flows driven in the interior, this last result looks anomalous until one realises that the interior dissipation contribution vanishes identically for a sphere \citep{Zhang01} and a spheroid \citep{Zhang04}. This is easily understood after realising that the inertial waves in these geometries are polynomials in (the cylindrical radius)  $s$ and $z$ (see the discussion on p118 of \cite{K93}) and partitioned into orthogonal vector spaces $\V_n$ where $n$ is the order of the polynomial (see p125 in \cite{K93}).  The interior dissipation integral is then
\beq
\int \bu_0^* \cdot \nabla^2 \bu_0\, dV = \frac{4}{\lambda^2} \int \bu_0^* \cdot  \frac{\partial ^2 \bu_0}{\partial z^2}\, dV = 0
\eeq
as $\partial ^2 \bu_0/\partial z^2 \in \V_{n-2}$ and hence orthogonal to $\bu_0 \in \V_n$. This explains why the dissipation doesn't vanish for a cylinder (inertial waves are not polynomials) but does also for an ellipsoidal (inertial waves are polynomials in $x$,  $y$ and $z$) - see \S5 of \cite{Vantieghem14}.

Figures \ref{ucsphere} and \ref{VelCorsphere} show the corrective viscous velocity defined in equation (\ref{renormalisation}) for the inertial mode (2,4,1). This is consistent with figure \ref{fig6} in showing that the diffusion-free boundary conditions lead to a smaller viscous corrective flow than stress-free boundary conditions. In particular, the stress-free boundary conditions give a corrective viscous velocity $O(E^{1/2})$
smaller than the nonslip case everywhere due to the interior viscous shear layers, while  the diffusion-free solution is fully $O(E)$ smaller. Appendix C shows that even higher order boundary conditions can be applied to reduce the decay rate still further.

%
%
%

\section{Inertial wave problem for a spherical shell}

\begin{figure}
\centering
\scalebox{0.7}[0.7]{\includegraphics{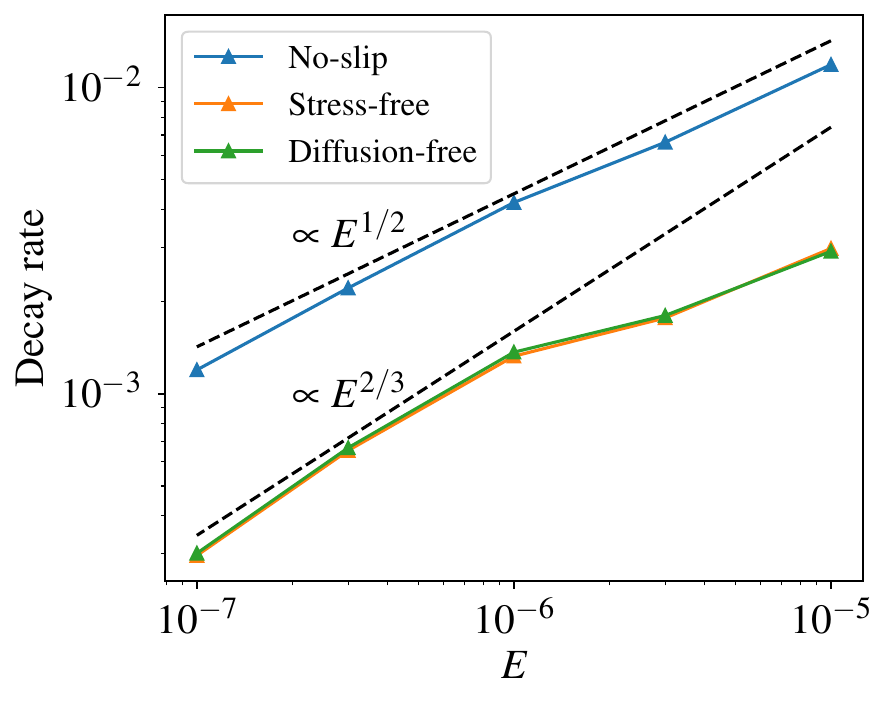}}
\caption{\label{shell} Decay rate of an inertial mode with $m=2$ and $\lambda=-1.4406$ in a spherical shell of radius ratio 0.5.}
\end{figure}

A spherical shell is an interesting singular geometry for the inertial wave problem as smooth inertial modes do not exist in the inviscid limit except for purely toroidal modes where the radial flow is zero everywhere. By adding a small viscosity, inertial waves exist in the form of conical shear layers spawned from the critical latitudes, though large scale structure may be hidden beneath localized shear layers at certain frequencies \citep{Lin21}.  Figure \ref{shell} shows the decay rate of an inertial wave with $m=2$ in a spherical shell of inner-to-outer radius ratio of 0.5. Now there is no obvious reduction in the corrective viscous velocity between the stress-free and diffusion-free boundary conditions. This is because there is no benefit of applying the latter if there is no solution of the inviscid problem. We would expect a benefit for the toroidal modes but have not checked this. 

%
%

%
%

\section{Discussion}

In this short note,  we have discussed an apparently little-known type of boundary condition that is a plausible alternative to the usual `stress-free' or Neumann conditions used as the boundary conditions which do least harm to the interior when diffusive processes are very weak.
Originally used in viscoelastic flow modelling,  the objective here has been to show how these  diffusion-free boundary conditions work for a couple of illustrative  (ODE)  model problems and then demonstrate that they work, as expected,  for a more complicated PDE problem.  The boundary layer correction is significantly weaker for the diffusion-free boundary conditions except when the diffusion-free problem doesn't have a well-defined solution itself (the spherical shell). Even in this latter scenario, diffusion-free does no worse than Neumann. We have also shown how these boundary conditions can help with hyperviscosity by acting as basis for generating as many boundary conditions as required.

Hopefully, these examples highlight how useful these boundary conditions could be in GAFD modelling.  The boundary conditions can either be phrased as `turning off the diffusion at the wall' or, equivalently for smooth solutions, imposing that the `diffusion term vanishes' which in its simplest form means setting the second derivative to zero.  The latter perspective is admittedly unusual: no textbooks discuss setting the  second derivative to vanish at a boundary for a second order equation.  Certainly for linear equations, this is because,  of course,  this is really just asking the linear combination of lower derivatives which define the second  order derivative to vanish which is a Robin condition. The point of the first model problem is, however, to show that this can be generalised to the nonlinear setting where the exact relationship is defined by turning diffusion off.

\section*{Acknowledgements}
RRK would like to thank Andrew Jackson and Greg Chini for  indirectly stimulating this note which was written for Andrew Soward's 80th birthday meeting held at Newcastle University in January 2024. 
The authors also thank Toby Wood for bringing the existence of the Ventcel/Ventsell/Wentzell boundary condition to their attention, Stephen K. Wilson for suggesting the `do least harm' moniker and Duncan Hewitt for asking about vanishing higher derivatives.

%
%

\appendices
%
%
\section{Extensions of ODE model}

Here we revisit the model problem in \S \ref{first_model} in order to examine the effect of using a) higher derivative boundary conditions and b) a higher order (hyper) diffusion term.

\subsection{Higher derivative boundary conditions}

It is straightforward to impose the boundary condition $y'''(1)=0$ or $y^{iv}(1)=0$ on the equation (\ref{model_1}) to get the leading order composite solutions
\beq
\begin{array}{ll}  
y = \displaystyle{\frac{1}{1+x}
+\frac{3}{8} \eps^3 e^{-(1-x)/\eps}  }         \hspace*{2cm}  
& {\rm for}\,\,\,{y'''(1)=0}, \\
 & \\
  y = \displaystyle{ \frac{1}{1+x} 
-\frac{3}{4}\eps^4 e^{-(1-x)/\eps}   }                  & {\rm for}\,\,\,{y^{iv}}(1)=0
\end{array}
\eeq
which therefore have $o(\eps^2)$ boundary layer corrections but the interior diffusive disturbance stays  $O(\eps^2)$. There is also another practical issue worth noting. While we have been able to impose the higher derivative boundary conditions directly, a numerical software package (e.g. Matlab's ODE15s) will expect these in terms of $y(1)$ and $y'(1)$ since equation (\ref{model_1}) is second order.  For the $y'''(1)=0$, this is straightforward to find by  differentiating $(\ref{model_1})$ and then using (\ref{model_1}) again to eliminate $y''(1)$ leaving
\beq
y'=\frac{ -y^2 }{ 1+2 \eps y }\biggl|_{x=1} \biggr. .
\eeq
Repeating this procedure, however, becomes more involved as the order of the boundary condition increases and can lead to difficulties. For example, $y^{iv}(1)=0$ leads to the boundary condition
\beq
2\eps y'^{\,2}+(1+4\eps y)y' +(1+2 \eps y)y^2=0|_{x=1}
\eeq
which is a quadratic for $y'$ and only one of the roots for $y'(1)$ in terms of $y(1)$ is equivalent to $y^{iv}(1)=0$.

\subsection{Hyperdiffusivity}

We now consider the model problem $y'=-y^2$ with a form of hyperviscosity
\beq
y'=-y^2-\E y^{iv}
\eeq
(the sign of the hyperviscosity is set by the need to have enough decaying solutions at $x=1$).
The boundary layer equation is
\beq
y_\xi=y_{\xi \xi \xi \xi}
\eeq
where $\xi:=(1-x)/\E^{1/3}$ with relevant solution
\beq
y= A \exp \biggl(\,\left[-\frac{1}{2}+i\frac{\sqrt{3}}{2} \right]\frac{1-x}{\E^{1/3}}\,\biggr)
+A^* \exp \biggl(\,\biggl[-\frac{1}{2}-i\frac{\sqrt{3}}{2} \biggr]\frac{1-x}{\E^{1/3}}\,\biggr)
\eeq
where $A^*$ is the complex conjugate of $A$. There are therefore two free (real) constants requiring two boundary conditions at $x=1$ (a further boundary condition is also needed at $x=0$). Just concentrating on those at $x=1$, we could impose $y'''(1)=y^{iv}(1)=0$
leading to $A=O(\E)$ or perhaps $y^{iv}(1)=y^{v}(1)=0$ in which case $A=O(\E^{4/3})$.

The point here is  that the order of the diffusion term ($n$ with $n=4$ here) sets the thickness scaling of the boundary layer as $O(\E^{1/(n-1)})$ and the lowest order derivative in the boundary conditions ($m$ with $m=3$ or $m=4$ here) sets the  order of the boundary layer correction at $O(\E^{m/(n-1)})$. In \S \ref{first_model}, diffusion-free b.c.s. have $m=2$ and ordinary (second order) diffusion $n=2$.

%
%
\section{Extensions of the inertial wave problem}

Here we examine the inertial wave problem in a cylinder with periodic boundary conditions applied along the axial direction to keep things simple. Boundary conditions then just need to be applied on the radial boundary $s=1$. 

\subsection{Higher derivative boundary conditions}

In \S \ref{cylinder}, non-slip, stress-free or diffusion-free boundary conditions were used which involve zero, first or second order normal (radial) derivatives of  the velocity field respectively. Here we show that higher (cylinder) radial derivatives can be used even though the diffusion is only second order in the radial variable. Specifically, we take
\beq 
u\,=\, \bhphi \cdot\partial_s^n (\Delta \bu) \,=\, \bhz \cdot \partial_s^n (\Delta\bu) \,=\, 0
\eeq
for some integer $n$ (the diffusion-free b.c.s have $n=0$). This is equivalent to imposing the usual inviscid b.c. $u=0$ at the sidewall $s=1$ and together with the  $n$th normal (radial) derivative of the tangential inviscid equations, i.e. 
\begin{align}
\frac{d^n}{ds^n}
\biggl[i \lambda v &+ 2u+\frac{imp}{s}\biggr] =0, \\
\frac{d^n}{ds^n}
\biggl[i \lambda w & \hspace{1.1cm}                        +i \alpha p \biggr] =0
\end{align}
where now azimuthal ($m$) and axial ($\alpha$) wavenumbers have been included. 
As in Appendix A, the boundary layer thickness stays $O(\sqrt{E})$ regardless of $n$ but the boundary layer correction becomes $O(E^{(n+2)/2})$ as its $(n+2)$th radial derivative must be $O(1)$ to compensate for the outer solution at the boundary. The boundary layer efflux sets the boundary layer contribution to the decay rate as $O(E^{(n+3)/2})$ which is subdominant to the interior's $O(E)$ contribution. Numerical calculations for $n=1$ readily confirm this:  at $E=10^{-4}$, $\lambda_i/E= 9.970798$ and at $E=10^{-6}$, $\lambda_i/E=9.970897$ where the interior contribution is $4 \alpha^2E/\lambda^2=9.970897E$ (for $m=1$ and $\alpha=\pi/1.9898$ used in the main text). So higher order boundary conditions can weaken the boundary layer correction as expected but this might be of limited impact because of the interior.

%
%
\begin{figure}
\centering
\scalebox{0.5}[0.5]{\includegraphics{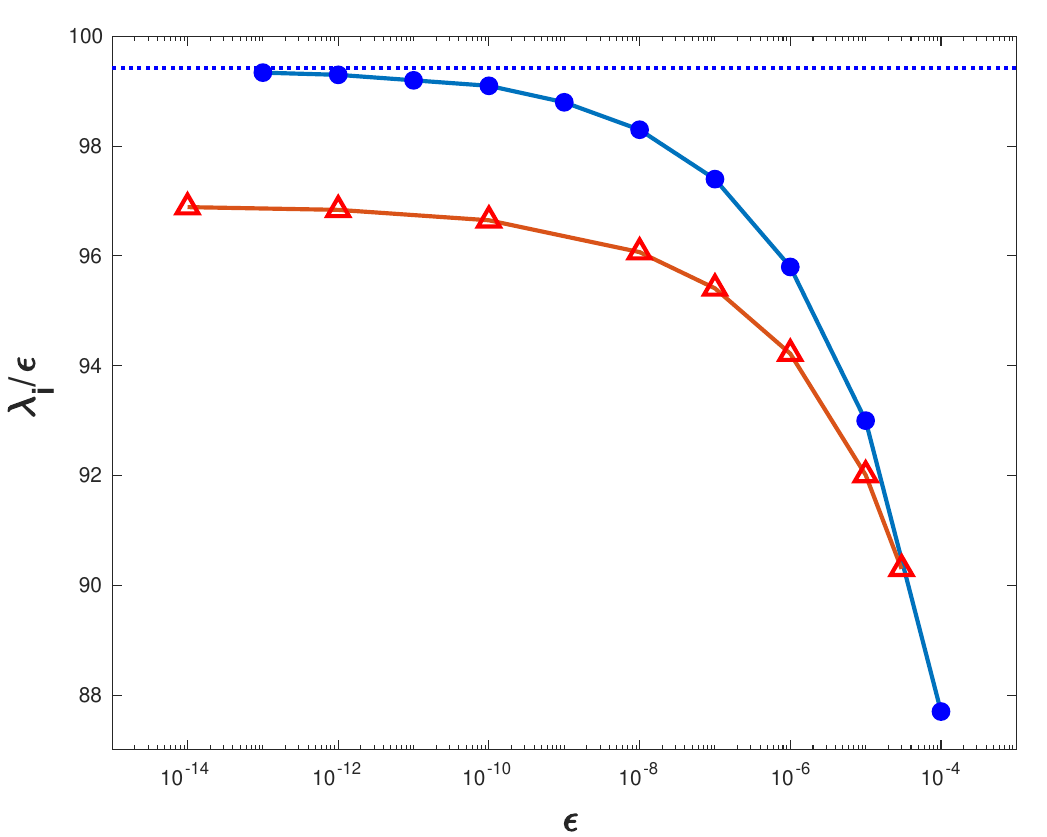}}
\caption{\label{higher} Decay rates $\lambda_i$ for case a) (blue solid circles) and case b) (red triangles) using hyperviscosity $-\E \Delta^2 {\bf u}$. The dashed line is drawn at $16\alpha^4/\lambda_r^4 \approx 99.4$ for $\alpha=\pi/1.9898$ and $\lambda_r \approx 1$ which is the expected interior contribution. This plot therefore indicates that the boundary layer contribution to the decay rate is  asymptotically subdominant (as $\E \rightarrow 0$) to that in the interior for case a) but is at the same order (and of opposite sign!) for case b).}
\end{figure}

\subsection{\label{hyper} Hyperdiffusivity}

Now we consider 
\beq
\frac{\partial \bu}{\partial t}+2 \bhz \times \bu +\bnab p=-\E \Delta^2\bu,    \label{1}
\eeq
\beq
\bnab . \bu=0, \label{2}
\eeq
as the simplest example of hyperviscosity where $\E \ll 1$ is a non-dimensionalised hyperviscosity parameter. The system now contains three 4th order equations and so requires 6 b.c.s at $s=1$ (the other 6 are regularity conditions at the axis). Two sets of boundary conditions are considered again in analogy to appendix A. They are
\begin{align}
& (a) \qquad u \, = \, \bhphi \cdot \Delta^2 \bu \,= \, \bhz \cdot \Delta^2 \bu \,= \,0  
\quad  \&  \quad 
\bhs \cdot \Delta^2 \bu \, = \, \bhphi \cdot \partial_s \Delta^2 \bu \,= \, \bhz \cdot \partial_s \Delta^2 \bu \,= \,0 \\
& (b) \qquad u \, = \, \bhphi \cdot \Delta^2 \bu \,= \, \bhz \cdot \Delta^2 \bu \,= \,0  
\quad  \&  \quad 
\bhs \cdot \Delta^2 \bu \, = \, \bhphi \cdot \partial_s \Delta \bu \,= \, \bhz \cdot \partial_s \Delta \bu \,= \,0 
\end{align}
The first 3 in both cases are  diffusion-free b.c.s as they require the tangential part of the hyperviscosity term to vanish on the boundary along with the inviscid, no-normal velocity condition. We, however, need 3 more and natural choices are forcing the normal component of the diffusion to vanish as well as either one higher (case (a)\,) or one lower (case (b)\,) normal boundary derivative of the tangential diffusion to vanish.
 In both cases the size of the boundary layer thickness is set by the order of hyperviscosity - here this is $O(\E^{1/4})$ - while the order of the boundary layer correction is set by the boundary condition with lower order wall-normal derivative. 
 In case (a) this will give a boundary layer contribution of $O(\E)$ and therefore a normal Ekman pumping velocity of $O(\E^{5/4})$ via incompressibility. Hence the boundary contribution to the decay rate will be $O(\E^{5/4})$ and so  subdominant to that of the interior which is straightforwardly derived as $16\alpha^4 \E/\lam^4 \approx 99.4\E$. This is consistent with numerical computations - see the blue line in figure \ref{higher} where the decay rate converges to the interior value as $\E \rightarrow 0$. 

In case (b), because the 3rd order derivative needs to be vanish, the boundary layer contribution is $O(\E^{3/4})$ and so the (hyper) Ekman pumping is now $O(\E)$. The boundary layer contribution is then $O(\E)$ which is the same order as the interior contribution. In this case the total decay rate will be $O(\E)$ but different from just the interior value - the red line in figure \ref{higher} clearly shows  this. Notice that the net effect of the boundary layer is to {\em reduce} the decay rate produced by the interior.

The conclusion here is then that diffusion-free boundary conditions also work well at weakening boundary layers for  hyperviscosity and reducing resolution requirements - see Table 4. The extra conditions also needed can be straightforwardly defined by either integrating or differentiating these  in the wall normal direction. The latter produces the weaker boundary layer correction but, since, it involves higher derivatives, requires more care to maintain numerical conditioning. 
Specifically, to generate figure \ref{higher}, new variables $\UU:=\bhs \cdot \Delta \bu$, $\VV:= \bhphi \cdot \Delta \bu$ and $\WW:= \bhz \cdot \Delta \bu$ were defined in  case (b) adding  3 extra equations to keep each to second order at most. Then the boundary conditions are
\beq
u =0 \qquad
i \lambda v + 2u+\frac{imp}{s} =0, \qquad
i \lambda w +i \alpha p =0,
\eeq
\beq
i \lambda u  -2v +\frac{dp}{ds} =0, \qquad 
\frac{d \VV}{ds} =0 \qquad
\frac{d \WW}{ds} =0
\eeq
on $s=1$. For case (a), we need to {\em further} define $\VVV=\bhphi \cdot \Delta \bUU$ and $\WWW:= \bhz \cdot \Delta \bUU$ where $\bUU:=\Delta \bu$ - so an extra 2  equations are added - and then the boundary conditions are 
\beq
u =0 \qquad
i \lambda v + 2u+\frac{imp}{s} =0, \qquad
i \lambda w +i \alpha p =0,
\eeq
\beq
i \lambda u  -2v +\frac{dp}{ds} =0, \qquad 
\frac{d \VVV}{ds} =0 \qquad
\frac{d \WWW}{ds} =0
\eeq
on $s=1$.

\begin{table}
\caption{
\label{case_a} Resolutions needed to resolve $\lambda$ in \S \ref{hyper} for case (a) with hyperviscosity as shown in figure \ref{higher}. $N$ is the number of Chebyshev polynomials used to represent each of the 10 variables $(u,v,w,p,\UU,\VV,\WW,\UUU,\VVV,\WWW)$ (so, for example, the matrices need only be $1000 \times 1000$ to handle $\E=10^{12}$). The inviscid frequency of the inertial wave is $1.000007648337$. 
}
\begin{center}
\begin{tabular}{@{}lrccc@{}}
&&&&\\
\hline\noalign{\smallskip}
&&&&\\
$\E$         & $N$ & \hspace{0.5cm}  &$\lambda_r$         & $\lambda_i/\E$ \\
&&&& \\ 
\hline\noalign{\smallskip}
&&&&\\
$10^{-4}$ & 50  && 0.999411036873915       &  -87.6892 \\ 
             & 75  && 0.999412672688673    &  -87.6968 \\
             &     &&                      &                    \\
$10^{-6}$     & 50  && 1.00000613867006    &  -95.8302  \\
             & 100 && 1.00000614172432     &  -95.8323 \\
             &     &&                     &                    \\             
$10^{-8}$ &  50 && 1.00000764363842       &  -98.2838 \\
             & 100 && 1.00000764364156    &  -98.2838 \\
             &     &&                     &                    \\
$10^{-10}$   & 100 &&  1.00000764832166   &   -99.0602\\
              & 150 &&  1.00000764832176   &   -99.0580 \\ 
             &     &&                     &                    \\
$10^{-12}$  & 100  &&  1.00000764833654  &     -99.2988 \\ 
             & 150  &&  1.00000764833651  &    -99.3042  \\
             &     &&                     &                    \\
\hline\noalign{\smallskip}            
\end{tabular}
\end{center}
\end{table}

%
%
\section{Toroidal-Poloidal scalar equations and boundary conditions in a sphere}

In a spherical geometry, we made use of the toroidal-poloidal decomposition (\ref{eq:TP}) and spherical harmonic expansions (\ref{eq:PP}), which transfer the inertial wave problem to a set of ODEs for the scalar functions $\TT_n(r)$ and $\PP_n(r)$. In this appendix, we show the projected equations and the corresponding boundary conditions. In addition, we show the effect of higher derivative boundary conditions for the decay rate of inertial waves in a sphere. 

Using eqs (\ref{eq:TP}-\ref{eq:PP}), the three components of the velocity $\bu$ can be expressed as 
\beqa
u_r &=& \sum_{n=m}^{N} n (n+1)\PP_n(r)Y_n^m(\theta,\phi), \\
u_\theta &=& \sum_{n=m}^{N}  \TT_n(r)\frac{\mathrm{i}m}{\sin \theta}Y_n^m(\theta,\phi)+\sum_{n=m}^{N}  \left[2\PP_n(r)+r \frac{d\PP_n(r)}{dr}\right]\frac{\partial}{\partial \theta}Y_n^m(\theta,\phi), \\
u_\phi &=& \sum_{n=m}^{N}  \TT_n(r)\frac{\partial}{\partial \theta}Y_n^m(\theta,\phi)+\sum_{n=m}^{N}  \left[2\PP_n(r)+r \frac{d\PP_n(r)}{dr}\right]\frac{\mathrm{i}m}{\sin \theta}Y_n^m(\theta,\phi).
\eeqa

 The diffusion term $\nabla^2 \bu$ can be expanded as
 \beqa
 (\nabla^2 \bu)_r &=& \sum_{n=m}^{N} n(n+1)\left[\frac{d^2\PP_n(r)}{dr^2}+\frac{4}{r}\frac{d\PP_n(r)}{dr}+(2-n-n^2)\frac{\PP_n(r)}{r^2} \right] Y_n^m(\theta,\phi), \\
  (\nabla^2 \bu)_\theta &=& \sum_{n=m}^{N}  \Gamma_n(r) \frac{\mathrm{i}m}{\sin \theta}Y_n^m(\theta,\phi)+\sum_{n=m}^{N}  \Pi_n(r) \frac{\partial}{\partial \theta}Y_n^m(\theta,\phi), \\
  (\nabla^2 \bu)_\phi &=&
 \sum_{n=m}^{N}  \Gamma_n(r) \frac{\partial}{\partial \theta}Y_n^m(\theta,\phi)+\sum_{n=m}^{N}  \Pi_n(r) \frac{\mathrm{i}m}{\sin \theta}Y_n^m(\theta,\phi),
 \eeqa
where
\beqa
\Gamma_n(r)&=& -\frac{d^2\TT_n(r)}{dr^2}-\frac{2}{r}\frac{d\TT_n(r)}{dr}+(n+n^2)\frac{\TT_n(r)}{r^2}, \\
\Pi_n(r)&=&r\frac{d^3\PP_n(r)}{dr^3}+6\frac{d^2\PP_n(r)}{dr^2}-(n^2+n-6)\frac{d\PP_n(r)}{rdr}.
\eeqa
Substituting the above formulas of $\bu$ and $\nabla^2 \bu$ into the inertial wave equation (\ref{eq:NS}), taking $\bR \cdot \bnabla \times$ and $\bR \cdot \bnabla \times \bnabla \times$ of the equation and using the orthogonality of spherical harmonics, we obtain the following equations:
 \begin{multline} \label{eq:TP_ode1}
 \mathrm{i}\lambda \TT_n-2\frac{\mathrm{i}m}{n(n+1)}\TT_n-2A_n^m\left[r\frac{d\PP_{n-1}}{dr}-(n-2)\PP_{n-1}\right]\\
 -2A_{n+1}^m\left[r\frac{d\PP_{n+1}}{dr}+(n+3)\PP_{n+1}\right]=E  \mathcal{D}_n^t(\TT_n),
 \end{multline}
 \begin{multline} \label{eq:TP_ode2}
 \mathrm{i}\lambda \mathcal{L}_n^p(\PP_n)-2\frac{\mathrm{i}m}{n(n+1)}\mathcal{L}_n^p(\PP_n)+2B_n^m\left[r\frac{d\TT_{n-1}}{dr}-(n-1)\TT_{n-1}\right]\\
 +2B_{n+1}^m\left[r\frac{d\TT_{n+1}}{dr}+(n+2)\TT_{n+1}\right]=E\mathcal{D}_n^p(\PP_n),
 \end{multline}
where 
\beqa
A_n^m&=&\frac{1}{n^2}\left(\frac{n^2-m^2}{4n^2-1}\right)^{1/2}, \\
B_n^m&=& n^2(n-1)^2A_n^m, \\
\mathcal{D}_n^t &=&\frac{d^2}{dr^2}+\frac{2}{r}\frac{d}{dr}-\frac{n(n+1)}{r^2},\\
\mathcal{L}_n^p &=&\frac{d^2}{dr^2}+\frac{4}{r}\frac{d}{dr}+\frac{2-n(n+1)}{r^2},\\
\mathcal{D}_n^p &=&r^2\frac{d^4}{dr^4}+8r\frac{d^3}{dr^3}-2(n^2+n-6)\frac{d^2}{dr^2}-\frac{4(n^2+n)}{r}\frac{d}{dr}+\frac{n(n^2-1)(n+2)}{r^2}.
\eeqa

\begin{figure}
\centering
\scalebox{0.6}[0.6]{\includegraphics{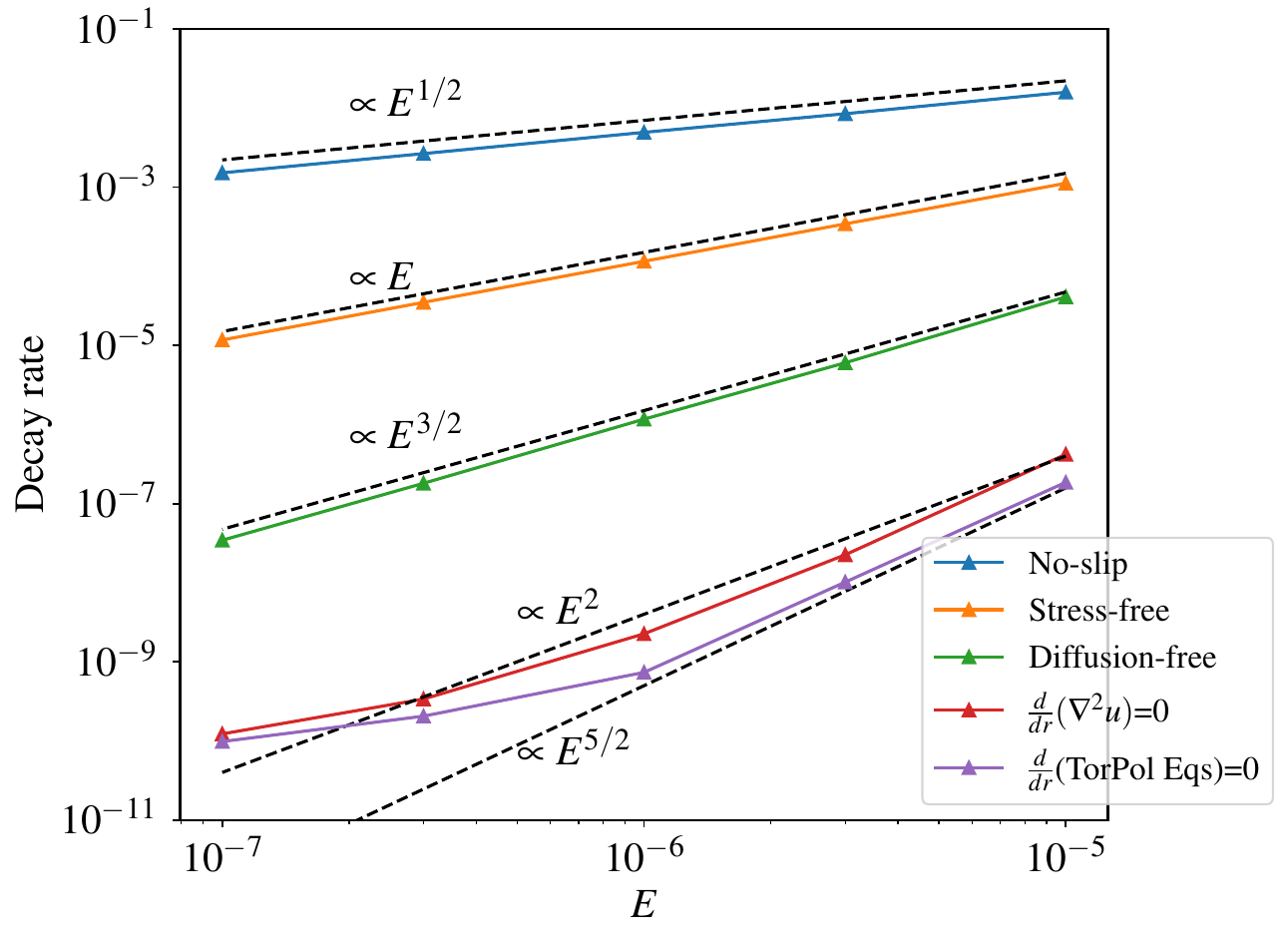}}
\caption{\label{highersphere} Decay rates of the inertial mode (2,4,1) in a sphere with higher derivative boundary conditions.}
\end{figure}

The diffusion-free boundary conditions (\ref{eq:dfbc_phy}) require
\beq
\PP_n(r)=0, \quad \Gamma_n(r)=0, \quad \Pi_n(r)=0,
\eeq
at the boundary $r=1$.
The higher radial derivative boundary conditions are  
\beq
u_r=\frac{d}{dr}(\nabla^2  \bu)_\theta=\frac{d}{dr}(\nabla^2  \bu)_\phi=0,
\eeq
which translate into 
\beq
\PP_n(r)=0, \quad \frac{d}{dr}\Gamma_n(r)=0, \quad \frac{d}{dr}\Pi_n(r)=0.
\eeq
The above diffusion-free boundary conditions involve the third radial derivative of the velocity $\bu$ and, using them, reduces the decay rate down to $O(E^{2})$ (the red curve in Figure \ref{highersphere}). 

We can also set the radial derivative of the viscous terms in the projected toroidal-poloidal equations (\ref{eq:TP_ode1}-\ref{eq:TP_ode2}) to zero as the boundary condition, i.e.
\beq
\PP_n(r)=0, \quad \frac{d}{dr}\left(\mathcal{D}_n^p(\PP_n)\right)=0, \quad \frac{d}{dr}\left(\mathcal{D}_n^t(\TT_n)\right)=0,
\eeq
which is equivalent to setting $\PP_n(r)=0$ and the radial derivative of the left hand side of equations (\ref{eq:TP_ode1}-\ref{eq:TP_ode2}) to zero at the boundary $r=1$. These boundary conditions involve four radial derivatives of the velocity $\bu$ and thereby lead to a further reduction of the decay rate to $O(E^{5/2})$ (the purple curve in Figure \ref{highersphere}).  The `turn-up' of both red and purple curves when $E \lesssim 10^{-6.5}$ seems caused by rounding numerical errors becoming important when trying to resolve decay rates $O(10^{-10})$ smaller than the frequency in double precision arithmetic. 


\section*{Disclosure statement}
No potential conflict of interest was reported by the authors.
\section*{Funding}
YL was supported by the National Natural Science Foundation of China (Nos. 12250012, 42142034) and RRK by the UK's EPSRC under grant EP/V027247/1.

\bibliographystyle{gGAF}

\vspace{36pt}

\markboth{Taylor \& Francis and I.T. Consultant}{Geophysical and Astrophysical Fluid Dynamics}


\end{document}